\documentclass[aps,prd,draft,nofootinbib,groupedaddress,preprintnumbers,%
              showpacs,showkeys]{revtex4}       % Documentstyle for
\usepackage{psfig,colordvi,amssymb,amsmath,float}
\usepackage[latin1]{inputenc}                    % Codepage latin1
\usepackage{graphicx} 
\usepackage{latexsym}                            % Load additional symbols
\usepackage{amsfonts}                            % Use AMS fonts,
\usepackage[mathscr]{eucal}                      % Euler font symbols
\usepackage{dcolumn}                             % Align table columns

\usepackage{slashed}
\usepackage{dsfont}                              % syntax for scale parsing

\usepackage{multirow}

\usepackage{bm}                                  % Bold math
\usepackage{hyperref}                            % References as links

\graphicspath{{.}{Graphics/}}                    % Path for graphics
\newcommand{\tr}{\mbox{Tr}}                      % Trace symbol

\newcommand{\be}{\begin{equation}}
\newcommand{\ee}{\end{equation}}
\newcommand{\bea}{\begin{eqnarray}}
\newcommand{\eea}{\end{eqnarray}}
\usepackage{slashed}
\usepackage{dsfont}
\usepackage[latin1]{inputenc}

\newcommand{\s}{{\mathcal S}} % Charge conjugation matrix
\newcommand{\Op}{\mathcal{O}} % Fractur O
\newcommand{\eins}{\mathds{1}} % Operator 1
\newcommand{\J}{\mathcal{J}} % operator 

\newcommand{\re}{\operatorname{Re}}

\newcommand{\hs}{\hspace{-1cm}}
\def\eq#1{\rm Eq.~(\ref{#1})}
\def\lsim{\mathrel{\rlap{\lower4pt\hbox{\hskip1pt$\sim$}}
    \raise1pt\hbox{$<$}}}                % less than or approx. symbol
\def\gsim{\mathrel{\rlap{\lower4pt\hbox{\hskip1pt$\sim$}}
    \raise1pt\hbox{$>$}}}                % greater than or approx. symbol

 \newcommand{\pslash}{{\not{\hspace{-0.08cm}p}}}  
   
 \newcommand{\csw}{\, c_{\rm SW}}  
   
\newcommand{\Dlr}{\buildrel \leftrightarrow \over D\raise-1pt\hbox{}}
 \newcommand{\Dl}{\buildrel \leftarrow \over D\raise-1pt\hbox{}}
\newcommand{\Dr}{\buildrel \rightarrow \over D\raise-1pt\hbox{}}

\begin{document}

% Setup title page
\preprint{} \title{Renormalization constants for 2-twist operators in twisted mass QCD}
\author{C.~Alexandrou~$^{a, b}$, M.~Constantinou~$^a$. T. Korzec~$^c$,
H.~Panagopoulos~$^a$, F. Stylianou~$^a$}
\affiliation{$^a$ {Department of Physics, University
  of Cyprus, PoB 20537, 1678 Nicosia, Cyprus \\
  $^b$  Computation-based Science and
  Technology Research Center, The Cyprus Institute, 
15 Kypranoros Str., 1645 Nicosia, Cyprus}\\
$^c$ {Institut f\"ur Physik,
   Humboldt Universit\"at zu Berlin, Newtonstrasse 15, 12489 Berlin,
   Germany} 
}
\email{{alexand@ucy.ac.cy},
  {marthac@ucy.ac.cy}, {korzec@physik.hu-berlin.de},
  {haris@ucy.ac.cy}, {fstyl01@ucy.ac.cy}} 
\date{\today}

\begin{abstract} 
Perturbative and
non-perturbative results on the renormalization constants  of the
fermion field and the twist-2 fermion
bilinears are presented with emphasis on the non-perturbative evaluation
 of the one-derivative twist-2 vector and axial
vector operators.
Non-perturbative results are obtained
using the twisted mass Wilson fermion formulation employing two degenerate
dynamical quarks and the tree-level Symanzik improved gluon action.
 The simulations have been performed for
pion masses in the range of about 450-260 MeV and at three values of
the lattice spacing $a$ corresponding to 
 $\beta=3.9,\,4.05,\,4.20$.
Subtraction of ${\cal O}(a^2)$ terms is
carried out by performing the perturbative evaluation of these
operators at 1-loop and up to ${\cal O}(a^2)$. 
The renormalization conditions are defined in the RI$'$-MOM
scheme, for both perturbative and non-perturbative results. The
renormalization factors, 
obtained for different values of the renormalization scale,
are evolved perturbatively to a reference scale set by the inverse of
the lattice spacing. In addition, they are translated to
${\overline{\rm MS}}$ at 2~GeV using 3-loop perturbative results for
the conversion factors.

\end{abstract}

\maketitle

%\begin{figure*}[h!]
%\centerline{\psfig{figure=ETMC.eps,height=2.2truecm}}
% \mbox{\epsfbox{ETMC.eps}}
%\end{figure*}

\newpage

\section{Introduction}
 Simulations in lattice QCD have advanced remarkably in the past couple of
 years reaching within 100 MeV of the physical pion mass.
This progress is due to theoretical improvements in defining the theory
on the lattice and to 
algorithmic improvements that give a better scaling behavior as the quark
mass decreases. These developments, combined with the tremendous increase in
computational power, have made {\it ab initio} calculations of
key observables on hadron structure in the chiral regime feasible enabling
comparison with experiment.
The hadron mass spectrum~\cite{Durr:2008zz, Alexandrou:2009qu} illustrates 
the good quality of lattice results that can be
obtained. The agreement with experiment is a validation of the lattice
approach, and justifies the computation of hadron observables beyond
hadron masses, such as form factors and parton distribution
functions. Both form factors and parton distribution functions can be
obtained from the so-called generalized parton distributions (GPDs) in
certain limiting cases. GPDs provide detailed information on the
internal structure of hadron in terms of both the longitudinal
momentum fraction and the total momentum transfer squared. Beyond the
information that the form factors yield, such as size, magnetization
and shape, GPDs encode additional information, relevant for experimental
investigations, such as the decomposition of the total hadron spin
into angular momentum and spin carried by quarks and gluons.

GPDs are single particle matrix elements of the light-cone
operator~\cite{Ji:1998pc, Hagler:2003jd},
\be
{\cal O}_{\Gamma}^f(x)=\int\frac{d\lambda}{4\pi}\> e^{i\lambda
    x}\,\overline{\psi}^f(-\frac{\lambda}{2} n)\,\Gamma \cdot n\, {\cal
    P}e^{ig\int_{-\lambda/2}^{\lambda/2} d\alpha \> n\cdot A(\alpha
    n)}\,\psi^f(\frac{\lambda}{2}n) \,  , 
\label{light-cone}
\ee
where $n$ is a light-cone vector, and ${\cal P}$ denotes a
path-ordering of the gauge fields in the exponential. Such matrix
elements cannot be calculated directly in lattice QCD. However,
${\cal O}(x)$ can be expanded in terms of local twist-two operators
\be {\cal O}_\Gamma^{f,\{\mu_1\mu_2\cdots\mu_n\}}= \overline{\psi}^f\Gamma^{\{\mu_1}i\Dlr^{\mu_2}\cdots i\Dlr^{\mu_n\}}\psi^f
\ee
 where
$\Dlr = \frac{1}{2}(\Dr - \Dl )$ and $\{\mu_1,\cdots,\mu_n\}$
denotes symmetrization of indices and subtraction of traces.
In this work we focus on the Dirac structures $\Gamma=\gamma^\mu,
 \gamma_5\,\gamma^\mu$ and $\gamma^5\,\sigma^{\mu\nu}$ 
($\sigma^{\mu\nu}=[\gamma^\mu,\gamma^\nu]/2$), which are referred to as
vector ${\cal O}_V^f(x)$, axial-vector ${\cal O}_A^f(x)$ and tensor
${\cal O}_T^f(x)$ operators, respectively. In lattice QCD we consider matrix elements
of such bilinear operators. A number of lattice groups are
producing results on nucleon form factors and first moments of
structure functions closer to the physical regime both in terms of
pion mass as well as in terms of the continuum limit~\cite{Hagler:2007xi,Syritsyn:2009mx,
Brommel:2007sb,Alexandrou:2008rp,Yamazaki:2009zq,Alexandrou:2009ng,
Alexandrou:tomasz}. While experiments are able to measure convolutions
of GPDs, lattice QCD allows us to extract hadron matrix elements for
the twist-2 operators, which can be expressed in terms of generalized
form factors. 
%\begin{equation}\label{eq_operators}
%           \Op_{V^a}^{\mu_1\ldots\mu_n} = \bar \psi\, D^{\{\mu_1}\cdots D^{\mu_{n-1}}\gamma^{\mu_n\}} \frac{\tau^a}{2} \, \psi,
%  \qquad  \Op_{A^a}^{\mu_1\ldots\mu_n} = \bar \psi\, D^{\{\mu_1}\cdots D^{\mu_{n-1}}\gamma^{\mu_n\}}\gamma^5 \frac{\tau^a}{2} \, \psi \, ,
%\end{equation}
%from which GPDs can be reconstructed via inverse Mellin transforms.

In order to compare hadron matrix elements of these local operators to
experiment one needs to renormalize them. The aim of this paper is to
calculate non-perturbatively the renormalization factors of the above
twist-two fermion operators within the twisted mass formulation. We
show that, although the lattice spacings considered in this work are
smaller than $1$~fm, ${\cal O}(a^2)$ terms are non-negligible and
significantly larger than statistical errors. We therefore compute the
${\cal O}(a^2)$-terms perturbatively and subtract them from the
non-perturbative results. This subtraction suppresses lattice
artifacts considerably depending on the operator under study and leads
to a more accurate determination of the renormalization constants.
Preliminary results of this work have been published in
Refs.~\cite{Alexandrou:tomasz,lat_martha}.

The paper is organized as follows: in Section \ref{sec2} we give the
expressions for the fermion and gluon actions we employed, and
define the twist-two operators. Section \ref{sec3} concentrates on the
perturbative procedure, and the ${\cal O}(a^2)$-corrected
expressions for the renormalization constants $Z_q$ and $Z_{\cal O}$. 
Section \ref{sec4} focuses on the
non-perturbative computation, where we explain the different steps of
the calculation. Moreover, we provide the renormalization
prescription of the RI$'$-MOM scheme, and we discuss alternative ways
for its application. 
The main results of this work are presented in Section
\ref{sec5}: the reader can find numerical values for the Z-factors of
the twist-2 operators, which are computed non-perturbatively and
corrected using the perturbative ${\cal O}(a^2)$ terms presented in
Section~\ref{sec3}. Since in general Z-factors depend on the
renormalization scale, we also provide results in the RI$'$-MOM scheme
at a reference scale, $\mu \sim 1/a$. For comparison with
phenomenological and experimental results, we convert the Z-factors to
the ${\overline{\rm MS}}$ scheme at 2~GeV. In Section~\ref{sec6} we
give our conclusions. 

A forthcoming paper \cite{Alexandrou:2010} will
focus on the perturbative procedure and will present results for local
fermion operators (scalar, pseudoscalar, vector, axial, tensor).

\section{Formulation}
\label{sec2}
\subsection{Lattice action}

For the gauge fields we use the tree-level Symanzik improved
gauge action~\cite{Weisz:1982zw}, which includes besides the
plaquette term $U^{1\times1}_{x,\mu,\nu}$ also rectangular $(1\times2)$ Wilson 
loops $U^{1\times2}_{x,\mu,\nu}$
\begin{equation}
    S_g =  \frac{\beta}{3}\sum_x\Biggl(  b_0\sum_{\substack{
      \mu,\nu=1\\1\leq\mu<\nu}}^4\left \{1-\re\tr(U^{1\times 1}_{x,\mu,\nu})\right \}\Bigr. 
     \Bigl.\,+
    \,b_1\sum_{\substack{\mu,\nu=1\\\mu\neq\nu}}^4\left \{1
    -\re\tr(U^{1\times2}_{x,\mu,\nu})\right \}\Biggr)\,  
\label{Symanzik}
\end{equation}
with $\beta=2\,N_c/g_0^2$, $b_1=-1/12$ and the
(proper) normalization condition $b_0=1-8b_1$. Note that at $b_1=0$ this
action becomes the usual Wilson plaquette gauge action.

The fermionic action for two degenerate flavors of quarks
 in twisted mass QCD is given by
\be
S_F= a^4\sum_x  \overline{\chi}(x)\bigl(D_W[U] + m_0 
+ i \mu_0 \gamma_5\tau^3  \bigr ) \chi(x)
\label{action}
\ee
with $\tau^3$ the Pauli matrix acting in
the isospin space, $\mu_0$ the bare twisted mass 
and $D_W$ the massless Wilson-Dirac operator defined as
\be
D_W[U] = \frac{1}{2} \gamma_{\mu}(\overrightarrow\nabla_{\mu} + \overrightarrow\nabla_{\mu}^{*})
-\frac{ar}{2} \overrightarrow\nabla_{\mu}
\overrightarrow\nabla^*_{\mu} 
\ee
where
\be
\overrightarrow\nabla_\mu \psi(x)= \frac{1}{a}\biggl[U_\mu(x)\psi(x+a\hat{\mu})-\psi(x)\biggr]
\hspace*{0.5cm} {\rm and}\hspace*{0.5cm} 
\overrightarrow\nabla^*_{\mu}\psi(x)=-\frac{1}{a}\biggl[U^\dagger_{\mu}(x-a\hat{\mu})\psi(x-a\hat{\mu})-\psi(x)\biggr]
\quad .
\ee
For completeness we also provide the definition of the backward
derivatives
\be
\overline\psi(x)\overleftarrow\nabla_\mu = \frac{1}{a}\biggl[\overline\psi(x+a\hat{\mu})U^\dagger_\mu(x)-\overline\psi(x)\biggr]
\hspace*{0.5cm} {\rm and}\hspace*{0.5cm} 
\overline\psi(x)\overleftarrow\nabla^*_{\mu}=-\frac{1}{a}\biggl[\overline\psi(x-a\hat{\mu})U_{\mu}(x-a\hat{\mu})-\overline\psi(x)\biggr]
\quad .
\ee
Maximally twisted Wilson quarks are obtained by setting the untwisted
bare quark mass $m_0$ to its critical value $m_{\rm cr}$, while the twisted
quark mass parameter $\mu_0$ is kept non-vanishing in order to give the
light quarks their mass. In $\eq{action}$ the quark fields $\chi$
are in the so-called ``twisted basis''. The ``physical basis'' is obtained for
maximal twist by the simple transformation
\be
\psi(x)=\exp\left(\frac {i\pi} 4\gamma_5\tau^3\right) \chi(x),\qquad
\overline\psi(x)=\overline\chi(x) \exp\left(\frac {i\pi} 4\gamma_5\tau^3\right)
\quad.
 \ee
In terms of the physical fields the action is given by
\be
S_F^{\psi}= a^4\sum_x  \overline\psi(x)\left(\frac 12 \gamma_\mu 
[\overrightarrow\nabla_\mu+\overrightarrow\nabla^*_\mu]-i \gamma_5\tau^3 \left(- 
\frac{ar}{2} \;\overrightarrow\nabla_\mu\overrightarrow\nabla^*_\mu+ m_{\rm cr}\right ) 
+  \mu_0 \right ) \psi(x)\quad.
\label{S_phy}
\ee

In this work we consider twist-two operators with one 
derivative, which are given in the twisted basis as follows
\begin{eqnarray}
\Op_{\rm DV}^{\{\mu\,\nu\}} &= \overline \chi \gamma_{\{\mu}\overleftrightarrow D_{\nu\}}\tau^a \chi 
                                              &= \begin{cases} \overline \psi  \gamma_5\gamma_{\{\mu}\overleftrightarrow D_{\nu\}} \tau^2 \psi   & a=1 \\
                                                              -\overline \psi  \gamma_5\gamma_{\{\mu}\overleftrightarrow D_{\nu\}} \tau^1 \psi   & a=2 \\
                                                               \overline \psi  \gamma_{\{\mu}\overleftrightarrow D_{\nu\}}         \tau^3 \psi   & a=3 \end{cases} \\[3ex]
\Op_{\rm DA}^{\{\mu\,\nu\}} &= \overline \chi \gamma_5\gamma_{\{\mu}\overleftrightarrow D_{\nu\}}\tau^a \chi 
                                              &= \begin{cases} \overline \psi  \gamma_{\{\mu}\overleftrightarrow D_{\nu\}} \tau^2 \psi   &\quad a=1 \\
                                                              -\overline \psi  \gamma_{\{\mu}\overleftrightarrow D_{\nu\}} \tau^1 \psi   & \quad a=2 \\
                                                               \overline \psi  \gamma_5\gamma_{\{\mu}\overleftrightarrow D_{\nu\}} \tau^3 \psi   & \quad a=3 \end{cases}\\[3ex]
\Op_{\rm DT}^{\mu\,\{\nu\,\rho\}} &= \overline \chi \gamma_5\sigma_{\mu\{\nu}\overleftrightarrow D_{\rho\}}\tau^a \chi 
                                              &= \begin{cases} \overline \psi  \gamma_5\sigma_{\mu\{\nu}\overleftrightarrow D_{\rho\}}\tau^a \psi   & a=1,2 \\
                         -i\,\overline \psi  \sigma_{\mu\{\nu}\overleftrightarrow D_{\rho\}}\eins \psi            & a=3 \end{cases}
\end{eqnarray}
with the covariant derivative defined as
\be
\Dlr = \frac{1}{2}\Big[\frac{(\overrightarrow\nabla_{\mu} +
    \overrightarrow\nabla_{\mu}^{*})}{2} -  \frac{(\overleftarrow\nabla_{\mu} +
    \overleftarrow\nabla_{\mu}^{*})}{2} \Big]\,.
\ee
The above operators are symmetrized over two Lorentz indices and are made traceless 
\be
\Op^{\{\sigma\,\tau\}} \equiv \frac{1}{2}\Big(\Op^{\sigma\,\tau}+\Op^{\tau\,\sigma}
\Big) - \frac{1}{4}\delta^{\sigma\,\tau} \sum_\lambda \Op^{\lambda\,\lambda}\,.
\ee
This definition avoids mixing with lower dimension operators.
We denote the corresponding Z-factors by $Z_{\rm DV}^a$, $Z_{\rm DA}^a$, $Z_{\rm DT}^a\,$.
In a massless renormalization scheme the renormalization constants are
defined in the chiral limit, where isospin symmetry is
exact. Hence, the same value for $Z$ is obtained
 independently of the value of the isospin
index $a$ and therefore we drop the $a$ index on the $Z$-factors from here
on. However, one must note that, for instance, the physical 
$\overline\psi \gamma_{\{\mu}\overleftrightarrow D_{\nu\}} \tau^1 \psi$ 
is renormalized with $Z_{\rm DA}$, while 
$\overline\psi \gamma_{\{\mu}\overleftrightarrow D_{\nu\}}\tau^3 \psi$
requires the $Z_{\rm DV}$, which differ from each other even in the chiral limit.

The one derivative operators fall into different irreducible
representations of the hypercubic group, depending on the choice of
indices. Hence, we distinguish between
\begin{eqnarray}
   \Op_{\rm DV1} &=& \Op_{\rm DV} \ {\rm with} \ \mu=\nu \\
\Op_{\rm DV2} &=& \Op_{\rm DV} \ {\rm with} \ \mu\neq\nu \\
   \Op_{\rm DA1} &=& \Op_{\rm DA} \ {\rm with} \ \mu=\nu \\
   \Op_{\rm DA2} &=& \Op_{\rm DA} \ {\rm with} \ \mu\neq\nu\\
   \Op_{\rm DT1} &=& \Op_{\rm DT} \ {\rm with} \ \mu\neq\nu=\rho\\
   \Op_{\rm DT2} &=& \Op_{\rm DT} \ {\rm with} \ \mu\neq\nu\neq\rho\neq\mu
\end{eqnarray}
Thus, $Z_{\rm DV1}$ will be different from $Z_{\rm DV2}$, but
renormalized matrix elements of the two corresponding operators will
be components of the same tensor in the continuum limit. $\Op_{\rm
  DT1}$ is sufficient to extract all generalized form factors of the
one derivative tensor operators. Although, in this work, we will
calculate non-perturbatively only the renormalization constants for
the vector and axial-vector operator, we will provide the perturbative
${\cal O}(a^2)$ terms also for the tensor operators.

\section{Perturbative procedure}
\label{sec3}

Here we present our results for the renormalization factor of the
fermion field, $Z_q$, that enters the evaluation of the twist-2
operators, and we also calculate $Z_{\rm DV},\,Z_{\rm DA},\,Z_{\rm DT}$. Our
calculation is performed in 1-loop perturbation theory to ${\cal
  O}(a^2)$. Extending the calculation up to ${\cal O}(a^2)$ brings in new
difficulties, compared to lowers order in $a$; for instance, there
appear new types of singularities. The procedure to address this issue
is extensively described in Ref.~\cite{Constantinou:2009tr}. 
Many IR singularities
encountered at ${\cal O}(a^2)$ would persist even up to 6 dimensions,
making their extraction more delicate. In addition to that, there
appear Lorentz non-invariant contributions in ${\cal O}(a^2)$ terms,
such as $\sum_\mu p_\mu^4/p^2$, where $p$ is the external momentum; as
a consequence, the Z-factors also depend on such terms.
The knowledge of the order $a^2$ terms is a big advantage for
non-perturbative estimates, since they can eliminate possible
large lattice
artifacts, once the ${\cal O}(a^2)$ perturbative terms are subtracted.

For all our perturbative results we employ a general fermion action,
which includes the clover parameter, $\csw$, and non-zero Lagrangian
mass, $m$. For the fermion field renormalization we also have a finite
twisted mass parameter, $\mu_0$, so we can explore its $\mu_0$ dependence.
Only the renormalization factors of the twist-2
operators are obtained at $\mu_0=0$, but we still consider $m\neq0$.
For gluons we use Symanzik improved actions (Plaquette, Tree-level
Symanzik, Iwasaki, TILW, DBW2) \cite{Constantinou:2009tr}. The purpose
of using such general fermion and gluon actions is to make our results
applicable to a variety of actions used nowadays in simulations.
The expressions for the matrix elements and the Z-factors are given in
a general covariant gauge, and their dependence on the coupling
constant, the external momentum, the masses and the clover parameter
is shown explicitly.

\subsection{Fermion field renormalization}

\begin{figure}[h]
\centerline{\psfig{figure=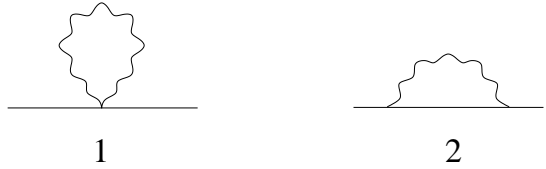,height=2.15truecm}}
\caption{One-loop diagrams contributing to the fermion
propagator. Wavy lines represent gluons and solid lines fermions.}
\label{fig1}
\end{figure}

The fermion action used for the inverse fermion propagator
improvement is a combination of Wilson/clover/twisted mass fermions, given by
\bea
\hspace{-0.75cm}
S_F &=& \sum_{x,\nu} \overline\psi(x)\Bigg[\frac{\gamma_\nu}{2}\left(\overrightarrow\nabla_\nu+\overrightarrow\nabla_\nu^\ast \right) -
\frac{a\,r}{2}\overrightarrow\nabla_\nu^\ast\,\overrightarrow\nabla_\nu + m_0 + i\,\mu_0\,\gamma_5\tau^3 
- \sum_\rho{1\over 4}\,c_{\rm SW}\,\sigma_{\rho\nu} {\hat F}_{\rho\nu}(x) \Bigg]\Psi(x) 
\label{action1}
\eea
All quantities appearing in Eq.~(\ref{action1}) are defined in Ref.~\cite{Constantinou:2009tr}.
There are two 1-loop diagrams involved in this particular computation,
which are illustrated in Fig. \ref{fig1}.

We compute $Z_q$ in the RI$'$-MOM renormalization scheme, defined as
\be
\hspace{-0.5cm}
Z_q = \frac{1}{4} {\rm Tr} \left[S^{(0)}_{\rm tree}(p)\,S^{-1}(p) \right]\Bigr|_{p_{\rho}=\mu_{\rho}} = 
-\frac{i}{4}{\rm Tr}\Bigg{[} 
\frac{\frac{1}{a}\sum_{\rho}\gamma_\rho \sin(a\,p_\rho)}{\frac{1}{a^2}
  \sum_{\rho} \sin^2(a\,p_\rho)}\,\cdot\,S^{-1}(p)\Bigg{]}_{p_{\rho}=\mu_{\rho}} \quad.
\ee
For comparison reasons, an additional renormalization prescription was
also applied (see Eqs.~(\ref{meth1}) - (\ref{meth1a})).
Since we want to take into account all ${\cal O}(a^2)$ terms, we
perform a Taylor expansion leading to 

\bea
Z_q = &-& \frac{i}{4}{\rm Tr}\Bigg{[} \frac{\sum_{\rho} \gamma_\rho ( p_\rho -\frac{a^2}{6}
      p^3_\rho)}{\sum_{\rho} p^2_\rho}\Bigg(1+\frac{a^2}{3}\frac{\sum_{\rho}
    p_\rho^4}{\sum_{\rho} p^2_\rho} \Bigg)
  \,\cdot\, S^{-1}_{\rm
    1-loop}(p)\Bigg{]}_{p_{\rho}=\mu_{\rho}} \hspace{-0.15cm} + {\cal O}(a^4\,g^2,g^4)\nonumber \\[3ex]
= &-& \frac{i}{4}{\rm Tr}\Bigg{[}
\frac{\pslash}{p^2} \,\cdot\, S^{-1}_{\rm 1-loop}(p) -
\frac{a^2}{3} \Bigg(\frac{1}{2} \frac{\pslash^3}{p^2} -
\frac{\pslash\,p4}{(p^2)^2} \Bigg)
\,\cdot\, S^{-1}_{\rm 1-loop}(p)
\Bigg{]}_{p_{\rho}=\mu_{\rho}} \hspace{-0.15cm} + {\cal O}(a^4\,g^2,g^4)\,\,.
\label{RC2}
\eea

\noindent The trace is taken only over spin indices and
$S^{-1}_{\rm{1-loop}}$ is the inverse fermion propagator that we
computed up to 1-loop and up to ${\cal O}(a^2)$. We make the following
definitions for convenience: $p^2\equiv\sum_\rho p^2_\rho$,
$p4\equiv\sum_\rho p^4_\rho$, $\pslash=\sum_{\rho}\gamma_\rho p_\rho$
and $\pslash^3\equiv\sum_{\rho}\gamma_\rho p^3_\rho$. A very important
observation is that the ${\cal O}(a^2)$ terms depend not only on the
magnitude, $\sum_\rho p^2_\rho$, but also on the direction of the
external momentum, $p_\rho$, as manifested by the presence of the
terms $\sum_\rho p^4_\rho$. As a consequence, alternative
renormalization prescriptions, involving different directions of the
renormalization scale $\mu_\rho=p_\rho$, treat lattice artifacts diversely. 

For the special choices:
$c_{SW}=0$, $r=1$ (Wilson parameter), $\lambda=0$ (Landau gauge),
$m_0=0$, $\mu_0=0$, and for tree-level Symanzik gluons, $Z_q$ can be
read from the following trace
%
% Z_q %
\bea
\frac{1}{4} {\rm Tr} \left[S^{(0)}_{\rm tree}(p)\,S^{-1}_{\rm 1-loop}(p) \right]\Bigr|_{p_{\rho}=\mu_{\rho}}
 &=& 1 + \tilde{g}^2 \Big{\{}{-}13.02327272(7) \nonumber \\[1ex]
&&\phantom{ 1 + \tilde{g}^2 \Big{\{}}
+ \Red{a^2} \Big[ \mu^2 \big( 1.14716212(5) -\frac{73}{360}
  \ln(\Red{a^2}\,\mu^2) \big) \nonumber \\[1ex]
&&\phantom{ 1 + \tilde{g}^2 \Big{\{} + \Red{a^2} }
{+}\frac{\sum_\rho \mu_\rho^4}{\mu^2} \big(2.1064995(2) -\frac{157}{180} \ln(\Red{a^2}\,\mu^2) \big)
\Bigr] \Bigr{\}} + {\cal O}(a^4\,g^2,g^4)
\eea
where $\tilde{g}^2 \equiv g^2C_F/(16\pi^2)$,
$C_F=(N_c^2-1)/(2N_c)$ and $\mu^2\equiv\sum_\rho \mu_\rho^2$.
In Ref.~\cite{Alexandrou:2010} we provide $Z_q$ for $\csw=0$,
$\lambda=0$ and the dependence on $m_0,\,\mu_0$ shown explicitly. Its
most general expression (including $\csw$ and $\lambda$, as well as a
wider choice of values for the parameters entering the Symanzik
action) is far too lengthy to be included in paper form ($Z_q$,
$Z_{\rm DV}$, $Z_{\rm DA}$, $Z_{\rm DT}$ around: 250, 800, 800, 950
terms respectively); it is provided in electronic form along with
Ref.~\cite{Alexandrou:2010}.

\vskip 1cm
\subsection{Renormalization of twist-2 operators}

Here we present the computation of the amputated Green's functions for
the following three twist-2 operators
\bea
{\cal O}^{\{\mu\nu\}}_{\rm DV} &=& \frac{1}{2}\Big[\overline\Psi\,\gamma_{\mu}\,\Dlr_{\nu}\,\Psi + 
\overline\Psi\,\gamma_{\nu}\,\Dlr_{\mu}\,\Psi \Big] 
-\frac{1}{4} \delta_{\mu\nu} \sum_\tau \overline\Psi\,\gamma_\tau\,\Dlr_{\tau}\,\Psi  \label{extV}\\ [2ex]
{\cal O}^{\{\mu\nu\}}_{\rm DA} &=& \frac{1}{2}\Big[\overline\Psi\,\gamma_5\gamma_{\mu}\,\Dlr_{\nu}\,\Psi + 
\overline\Psi\,\gamma_5\gamma_{\nu}\,\Dlr_{\mu}\,\Psi \Big] 
-\frac{1}{4} \delta_{\mu\nu} \sum_\tau \overline\Psi\,\gamma_5\gamma_\tau\,\Dlr_{\tau}\,\Psi \label{extA}\\ [2ex]
{\cal O}^{\mu\{\nu\rho\}}_{\rm DT} &=& \frac{1}{2}\Big[\overline\Psi\,\gamma_5\sigma_{\mu\nu}\,\Dlr_{\rho}\,\Psi + 
\overline\Psi\,\gamma_5\sigma_{\mu\rho}\,\Dlr_{\nu}\,\Psi \Big] 
-\frac{1}{4} \delta_{\nu\rho} \sum_\tau \overline\Psi\,\gamma_5\sigma_{\mu\tau}\,\Dlr_{\tau}\,\Psi 
\label{extT}
\eea
which, being symmetrized and traceless, have no mixing with lower
dimension operators. 

The Feynman diagrams that enter our calculation are given below, where
the insertion of the twist-2 operator is represented by a cross.
\begin{figure}[h]
\centerline{\psfig{figure=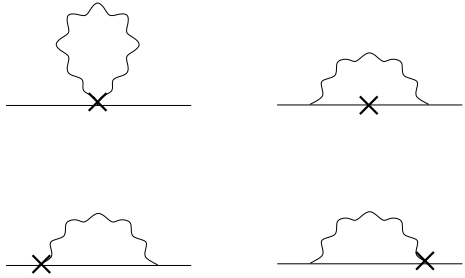,height=3.65truecm}}
\caption{One-loop diagrams contributing to the computation of the
twist-2 operators. A wavy (solid) line represents gluons
(fermions). A cross denotes an insertion of the operator under study.}
\label{fig2}
\end{figure}
We have computed, to ${\cal O}(a^2)$, the forward matrix
elements of these operators for general external indices $\mu,\,\nu$
(and $\rho$ for the tensor operator), external momentum $p,\,m_0,\,g,\,N_c,\,a,\,\csw$
and $\lambda$. Setting $\lambda=1$ corresponds to the Feynman gauge,
whereas $\lambda=0$ corresponds to the Landau gauge. Our final results
were obtained for the 10 sets of Symanzik coefficients we have used in the
calculation of $Z_q$ \cite{Constantinou:2009tr}.

In order to define $Z_{\cal O}$, we have used a renormalization
prescription which is most amenable to non-perturbative treatment:
\be
Z_q^{-1} Z_{O}{\rm Tr}\Big{[} L^{O}(p)\,\cdot\,L^{O}_{\rm tree}(p)\Big{]}_{p_{\lambda}=\mu_{\lambda}} = 
{\rm Tr}\Big{[}L^{O}_{\rm tree}(p) \,\cdot\,L^{O}_{\rm tree}(p)\Big{]}_{p_{\lambda}=\mu_{\lambda}}
\label{ZO}
\ee
where $L^{O}$ denotes the amputated 2-point Green's function of the
operators in Eqs.~(\ref{extV}) - (\ref{extT}), up to 1-loop and up to ${\cal O}(a^2)$. 
The tree-level expressions of the operators including the ${\cal O}
(a^2)$ terms are 
\bea
L^{\rm DV1}_{\rm tree}(p) &=& i \gamma_\mu \Big(p_\mu -
\Red{a^2}\,\frac{p_\mu^3}{6}\Big) - \frac{i}{4} \sum_\tau \gamma_\tau \Big(p_\tau -
\Red{a^2}\,\frac{p_\tau^3}{6}\Big)+ {\cal O}(a^4)\,,\quad L^{\rm DA1}_{\rm tree}(p) =
\gamma_5\,L^{\rm DV1}_{\rm tree}(p) \\ [1.5ex]
L^{\rm DV2}_{\rm tree}(p) &=& \frac{i}{2}\,\left( \gamma_\mu \Big(p_\nu -
\Red{a^2}\,\frac{p_\nu^3}{6}\Big) + \gamma_\nu \Big(p_\mu - \Red{a^2}\,\frac{p_\mu^3}{6}\Big)\right)+ {\cal O}(a^4)\,,\quad L^{\rm DA2}_{\rm tree}(p) =
\gamma_5\,L^{\rm DV2}_{\rm tree}(p)\\ [1.5ex]
L^{\rm DT1}_{\rm tree}(p) &=& i \gamma_5\,\sigma_{\mu\nu} \Big(p_\nu -
\Red{a^2}\,\frac{p_\nu^3}{6}\Big) - \frac{i}{4} \sum_\tau \gamma_5\,\sigma_{\mu\tau} \Big(p_\tau -
\Red{a^2}\,\frac{p_\tau^3}{6}\Big)+ {\cal O}(a^4)\\ [1.5ex]
L^{\rm DT2}_{\rm tree}(p) &=& \frac{i}{2}\,\gamma_5\,\left( \sigma_{\mu\nu} \Big(p_\rho -
\Red{a^2}\,\frac{p_\rho^3}{6}\Big) + \sigma_{\mu\rho} \Big(p_\nu - \Red{a^2}\,\frac{p_\nu^3}{6}\Big)\right)+ {\cal O}(a^4)
\eea
We perform a Taylor expansion up to ${\cal O}(a^2)$ in the right hand
side of the renormalization condition and it leads to the following
\bea
\hs {\rm Tr}\Big{[} L^{\rm DV1}_{\rm tree}(p)\,\cdot\,L^{\rm DV1}_{\rm tree}(p)\Big{]} &=& 
-2\,p_{\mu}^2 -\frac{1}{4} p^2 + \Red{a^2} (\frac{1}{12} p4 + \frac{2}{3} p_{\mu}^4) 
+ {\cal O}(a^4)\,=\,-  {\rm Tr}\Big{[} L^{\rm DA1}_{\rm
    tree}(p)\,\cdot\,L^{\rm DA1}_{\rm tree}(p)\Big{]} \\ [3ex]
\hs {\rm Tr}\Big{[} L^{\rm DV2}_{\rm tree}(p)\,\cdot\,L^{\rm DV2}_{\rm tree}(p)\Big{]} &=& 
-p_{\mu}^2 - p_\nu^2 + \frac{\Red{a^2}}{3} (p_{\mu}^4 + p_\nu^4) +
{\cal O}(a^4) \,=\,-  {\rm Tr}\Big{[} L^{\rm DA2}_{\rm
    tree}(p)\,\cdot\,L^{\rm DA2}_{\rm tree}(p)\Big{]}\\ [3ex] 
\hs {\rm Tr}\Big{[} L^{\rm DT1}_{\rm tree}(p)\,\cdot\,L^{\rm DT1}_{\rm tree}(p)\Big{]} &=& 
\frac{p^2}{4}  + 2\,p_\nu^2 - \frac{p_\mu^2}{4}  
- \Red{a^2} \left( \frac{p4}{12} +  \frac{2\,p_\nu^4}{3} -
\frac{p_\mu^4}{12}\right) + {\cal O}(a^4) \\ [3ex] 
\hs {\rm Tr}\Big{[} L^{\rm DT2}_{\rm tree}(p)\,\cdot\,L^{\rm DT2}_{\rm tree}(p)\Big{]} &=& 
p_\nu^2 + p_\rho^2 - \frac{\Red{a^2}}{3} \left(p_\nu^4 +
  p_\rho^4\right)  + {\cal O}(a^4) 
\eea
For the special choices: $m_0=0,\,c_{SW}=0$, $r=1$, $\lambda=0$ (Landau
gauge), and for tree-level Symanzik gluons, we obtain for the left
hand side of Eq.~(\ref{ZO})
{\small{
\bea
\label{LDV1}
\hs {\rm Tr}\Big{[} L^{\rm DV1}(p)\cdot L^{\rm DV1}_{\rm tree}(p)\Big{]} &=& 
-2\,p_{\mu}^2 -\frac{1}{4} p^2 + \Red{a^2} (\frac{1}{12} p4 + \frac{2}{3} p_{\mu}^4) \nonumber \\
&+& \tilde{g}^2 \Big{\{}
\frac{4}{3} \frac{p_{\mu}^4}{p^2} 
+ p^2 \, \big (3.610062(3) - \frac{2}{3} \ln(\Red{a^2}\,p^2) \big)
+ p_{\mu}^2 \, \big (27.54716(3) - \frac{16}{3} \ln(\Red{a^2}\,p^2) \big) \nonumber \\
&&\phantom{\tilde{g}^2 \Big{\{}\,\,} 
+\Red{a^2} \Big[ 
(p^2)^2 \, \big ( 0.11838(2) + \frac{7}{288} \ln(\Red{a^2}\,p^2) \big ) 
+p^2\,p_{\mu}^2 \, \big ({-}0.6573(1) - \frac{299}{180} \ln(\Red{a^2}\,p^2) \big )\nonumber \\
&&\phantom{\tilde{g}^2 \Big{\{}+\Red{a^2} \Big[} 
+p4 \, \big ({-}1.71886(3) + \frac{397}{720} \ln(\Red{a^2}\,p^2) - \frac{43}{360} \frac{p_{\mu}^2}{p^2} \big )\nonumber \\
&&\phantom{\tilde{g}^2 \Big{\{}+\Red{a^2} \Big[} 
+p_{\mu}^4\,\big ({-}16.1049(5) + \frac{94}{15} \ln(\Red{a^2}\,p^2) + \frac{29}{90} \frac{p4}{(p^2)^2} + \frac{169}{45} \frac{p_{\mu}^2}{p^2} \big )
\Bigr] 
\Bigr{\}}  \nonumber \\
&+& {\cal O}(a^4,g^4)
\eea
%
% n1NEQn2 V %
\bea
\hs {\rm Tr}\Big{[} L^{\rm DV2}(p)\cdot L^{\rm DV2}_{\rm tree}(p)\Big{]} &=& 
-p_{\mu}^2 - p_{\nu}^2 + \frac{\Red{a^2}}{3} (p_{\mu}^4 + p_{\nu}^4) \nonumber \\
&+& \tilde{g}^2 \Big{\{}
\frac{4}{3} \frac{p_{\mu}^2\,p_{\nu}^2}{p^2} 
+ (p_{\mu}^2+p_{\nu}^2) \, \big (15.04575(1) - \frac{8}{3} \ln(\Red{a^2}\,p^2) \big)\nonumber \\
&&\phantom{\tilde{g}^2 \Big{\{}\,\,} 
+\Red{a^2} \Big[ 
(p_{\mu}^4 + p_{\nu}^4 ) \, \big ({-}7.1429(1) + \frac{491}{360} \ln(\Red{a^2}\,p^2) \big ) \nonumber \\
&&\phantom{\tilde{g}^2 \Big{\{}+\Red{a^2} \Big[} 
+ (p_{\mu}^2 + p_{\nu}^2 ) \, \Big (p^2\,\big({-}0.13212(3) - \frac{103}{360} \ln(\Red{a^2}\,p^2) \big) + \frac{353}{720} \frac{p4}{p^2} \Big ) \nonumber \\
&&\phantom{\tilde{g}^2 \Big{\{}+\Red{a^2} \Big[} 
+p_{\mu}^2 \,p_{\nu}^2\,\big(
-4.0096(1)
+ \frac{1013}{180} \ln(\Red{a^2}\,p^2)
+ \frac{29}{90} \frac{p4}{(p^2)^2}
+\frac{169}{90} \frac{(p_{\mu}^2+p_{\mu}^2)}{p^2} 
\big)
\Bigr] 
\Bigr{\}}  \nonumber \\
&+& {\cal O}(a^4,g^4) 
\eea
%
% n1EQn2 A % 
\bea
\hs {\rm Tr}\Big{[} L^{\rm DA1}(p)\cdot L^{\rm DA1}_{\rm tree}(p)\Big{]} &=& 
2\,p_{\mu}^2 +\frac{1}{4} p^2 + \Red{a^2} ({-}\frac{1}{12} p4 - \frac{2}{3} p_{\mu}^4) \nonumber \\
&+& \tilde{g}^2 \Big{\{}
{-}\frac{4}{3} \frac{p_{\mu}^4}{p^2} 
+ p^2 \, \big ({-}4.127332(3) + \frac{2}{3} \ln(\Red{a^2}\,p^2) \big)
+ p_{\mu}^2 \, \big ({-}31.68532(3) + \frac{16}{3} \ln(\Red{a^2}\,p^2) \big) \nonumber \\
&&\phantom{\tilde{g}^2 \Big{\{}\,\,} 
+\Red{a^2} \Big[ 
(p^2)^2 \, \big ( 0.17035(2) + \frac{65}{288} \ln(\Red{a^2}\,p^2) \big ) 
+p^2\,p_{\mu}^2 \, \big (0.3982(1) - \frac{541}{180} \ln(\Red{a^2}\,p^2) \big )\nonumber \\
&&\phantom{\tilde{g}^2 \Big{\{}+\Red{a^2} \Big[} 
+p4 \, \big (1.69230(3) - \frac{397}{720} \ln(\Red{a^2}\,p^2) + \frac{43}{360} \frac{p_{\mu}^2}{p^2} \big )\nonumber \\
&&\phantom{\tilde{g}^2 \Big{\{}+\Red{a^2} \Big[} 
+p_{\mu}^4\,\big (18.4613(5) + \frac{2}{5} \ln(\Red{a^2}\,p^2) - \frac{29}{90} \frac{p4}{(p^2)^2} - \frac{169}{45} \frac{p_{\mu}^2}{p^2} \big )
\Bigr] 
\Bigr{\}}  \nonumber \\
&+& {\cal O}(a^4,g^4)
\eea
%
% n1NEQn2 A % 
\bea
\hs {\rm Tr}\Big{[} L^{\rm DA2}(p)\cdot L^{\rm DA2}_{\rm tree}(p)\Big{]} &=& 
p_{\mu}^2 + p_{\nu}^2 - \frac{\Red{a^2}}{3} (p_{\mu}^4 + p_{\nu}^4) \nonumber \\
&+& \tilde{g}^2 \Big{\{}
{-}\frac{4}{3} \frac{p_{\mu}^2\,p_{\nu}^2}{p^2} 
+ (p_{\mu}^2+p_{\nu}^2) \, \big (-16.10196(1) + \frac{8}{3} \ln(\Red{a^2}\,p^2) \big)\nonumber \\
&&\phantom{\tilde{g}^2 \Big{\{}\,\,} 
+\Red{a^2} \Big[ 
(p_{\mu}^4 + p_{\nu}^4 ) \, \big (7.2286(1) - \frac{491}{360} \ln(\Red{a^2}\,p^2) \big ) \nonumber \\
&&\phantom{\tilde{g}^2 \Big{\{}+\Red{a^2} \Big[} 
+ (p_{\mu}^2 + p_{\nu}^2 ) \, \Big (p^2\,\big(0.75869(3) - \frac{137}{360} \ln(\Red{a^2}\,p^2) \big) - \frac{353}{720} \frac{p4}{p^2} \Big ) \nonumber \\
&&\phantom{\tilde{g}^2 \Big{\{}+\Red{a^2} \Big[} 
+p_{\mu}^2 \,p_{\nu}^2\,\big(
4.8509(1)
+ \frac{187}{180} \ln(\Red{a^2}\,p^2)
- \frac{29}{90} \frac{p4}{(p^2)^2}
-\frac{169}{90} \frac{(p_{\mu}^2+p_{\mu}^2)}{p^2} 
\big)
\Bigr] 
\Bigr{\}}  \nonumber \\
&+& {\cal O}(a^4,g^4)\label{LDA2} 
\eea
\bea
\hs {\rm Tr}\Big{[} L^{\rm DT1}(p)\cdot L^{\rm DT1}_{\rm tree}(p)\Big{]} &=&
\frac{p^2-p_\mu^2}{4} + 2\,p_\nu^2 + 
   \Red{a^2}\,(\frac{p_\mu^4-p4}{12} - \frac{2\,p_\nu^4}{3}) \nonumber \\
&+&\tilde{g}^2\,\Big{\{} (p_\mu^2 -p^2)\,\big(4.226559(3) -  \ln(\Red{a^2}\,p^2) \big)  + 
      p_\nu^2 \big(-29.11666(2) \,+ 5\, \ln(\Red{a^2}\,p^2) \big)\nonumber \\
&&\phantom{\tilde{g}^2 \Big{\{}\,\,} 
 +\Red{a^2}\,\Big[ p^4 \,\big(-0.14754(2)  -  \frac{43}{1440} \, \ln(\Red{a^2}\,p^2)  \big)\nonumber \\ 
&&\phantom{\tilde{g}^2 \Big{\{}\,\,} 
+ p4  \,\big(1.93789(3) -  \frac{433}{720} \, \ln(\Red{a^2}\,p^2) - \frac{379}{720}\,\frac{p_\nu^2}{p^2}+ 
       \frac{17}{192}\,\frac{p_\mu^2}{p^2} \big)\nonumber \\ 
&&\phantom{\tilde{g}^2 \Big{\{}\,\,} 
+ p^2 \, \big(1.7215(1) \,p_\nu^2 +  \frac{61}{48} \, \ln(\Red{a^2}\,p^2)\,p_\nu^2 
+  0.37022(2) \,p_\mu^2  - \frac{227}{1440} \, \ln(\Red{a^2}\,p^2) \,p_\mu^2  \big)\nonumber \\ 
&&\phantom{\tilde{g}^2 \Big{\{}\,\,} 
+ p_\nu^4 \, \big(14.9155(4) -  \frac{71}{15} \, \ln(\Red{a^2}\,p^2) - 
       \frac{721}{90}\,\frac{p_\mu^2}{p^2}  \big)\nonumber \\ 
&&\phantom{\tilde{g}^2 \Big{\{}\,\,} 
+ p_\nu^2 \,p_\mu^2  \,\big(2.4896(1)   -  \frac{881}{240} \, \ln(\Red{a^2}\,p^2)  
-\frac{39}{10}\,\frac{p_\mu^2}{p^2} \big) \nonumber \\ 
&&\phantom{\tilde{g}^2 \Big{\{}\,\,} 
+ p_\mu^4 \, \big(-2.24911(4)+  \frac{71}{90} \, \ln(\Red{a^2}\,p^2)\big) 
- \frac{134}{45}\,\frac{p_\nu^6}{p^2}  \Big]\Bigr{\}}  \nonumber \\
&+& {\cal O}(a^4,g^4)
\eea
\bea
\hs {\rm Tr}\Big{[} L^{\rm DT2}(p)\cdot L^{\rm DT2}_{\rm tree}(p)\Big{]} &=&
 p_\rho^2 + p_\nu^2 + \Red{a^2}\,(-\frac{p_\rho^4}{3} - \frac{p_\nu^4}{3}) \nonumber \\
&+& \tilde{g}^2\,\Big{\{} (p_\nu^2 + p_\rho^2)\,\big(-15.84740(1) + 3\, \ln(\Red{a^2}\,p^2) \big) \nonumber \\
&&\phantom{\tilde{g}^2 \Big{\{}\,\,} 
+ \Red{a^2}\,\Big[(p_\nu^2 + p_\rho^2)\, \big(0.22134(3) \,p^2  +  \frac{107}{360} \, \ln(\Red{a^2}\,p^2) \,p^2 
- \frac{41}{60}\, \frac{p4}{p^2}  \nonumber \\
&&\phantom{\tilde{g}^2 \Big{\{}\,\,a^2\,\Big[(p_\nu^2 + p_\rho^2)\,\,\,\,\,\,\,} +  0.73604(2)\,p_\mu^2 
- \frac{301}{360} \, \ln(\Red{a^2}\,p^2) \,p_\mu^2 
- \frac{67}{90} \,\frac{p_\mu^4}{p^2} \big) 
   \nonumber \\ 
&&\phantom{\tilde{g}^2 \Big{\{}\,\,} 
- \frac{67}{15} \,\frac{p_\rho^2\,p_\mu^2\,p_\nu^2}{p^2} 
+ (p_\nu^4 + p_\rho^4)\, \big( 7.3949(1) -  \frac{1051}{720} \, \ln(\Red{a^2}\,p^2) \big)\nonumber \\ 
&&\phantom{\tilde{g}^2 \Big{\{}\,\,} 
 + p_\rho^2\,p_\nu^2 \big( 2.98450(8) \,    -  \frac{1609}{360} \, \ln(\Red{a^2}\,p^2) \big)
- \frac{67}{45}\,\frac{p_\rho^4\,p_\nu^2+ p_\rho^2\,p_\nu^4}{p^2} \Big]\Bigr{\}}  \nonumber \\
&+& {\cal O}(a^4,g^4)
\eea
}}
In our forthcoming publication~\cite{Alexandrou:2010}, we will include an ASCII file with all
our numerical results for general value of $\lambda,\,\csw,\,m_0$ and
the 10 sets of Symanzik gluon actions; the file is best perused as
Mathematica input. In addition, there appears an Appendix providing
the exact ${\cal O}(a^2)$ terms that need to be subtracted from $Z_O$
(a fraction of the two traces and  $Z_q$ of Eq.~(\ref{ZO})).

The ${\cal O}(a^2)$ terms shown in Eqs.~(\ref{LDV1}) - (\ref{LDA2}) are
used to correct our non-perturbative results for $Z_{\rm DV1}$, $Z_{\rm
  DV2}$, $Z_{\rm DA1}$, $Z_{\rm DA2}$, in order to better control $a^2$
artifacts. For the subtraction procedure we use the boosted coupling
\cite{Lepage:1992xa} instead of the bare one
\be
g^2_{\rm boosted} = \frac{g^2_{\rm bare}}{\langle u_{plaq} \rangle}\,\,,
\ee
where $\langle u_{plaq} \rangle$ is the plaquette mean value.

\section{Non-perturbative calculation}
\label{sec4}

\subsection{Evaluation of correlators}
In the literature there are two main approaches
that have been employed for the non-perturbative evaluation of the renormalization 
constants. They both start by considering that 
the operators can all be written in the form
\begin{equation}
   \Op(z) = \sum_{z'} \overline u(z) \J(z,z') d(z')\, ,
\end{equation}
where $u$ and $d$ denote quark fields in the physical basis and $\J$ denotes the operator we are interested in,
e.g. $\J(z,z') = \delta_{z,z'} \gamma_\mu$ would correspond to the local vector current.
For each operator we define a bare vertex function given by
\begin{equation}\label{vfun}  
   G(p) = \frac{a^{12}}{V}\sum_{x,y,z,z'} e^{-ip(x-y)} \langle u(x) \overline u(z) \J(z,z') d(z') \overline d(y) \rangle \, ,
\end{equation}
where $p$ is a momentum allowed by the boundary conditions, $V$ is the lattice volume, and the gauge average is
performed over gauge-fixed configurations. We have suppressed the Dirac and color indices 
of $G(p)$.
The first approach relies on translation invariance to shift the coordinates
of the correlators in
Eq.~(\ref{vfun}) to position $z=0$ \cite{Dimopoulos:2007fn, Zhaofeng}. 
Having shifted to $z=0$ allows one to calculate the 
amputated vertex function for a given operator $\J$ for {\it any} momentum
with one inversion per quark flavor.

In this work we explore the second approach, 
introduced in Ref.~\cite{Gockeler:1998ye},
which uses directly Eq.~(\ref{vfun}) without employing translation
invariance. One must now use a source that is momentum dependent but
can couple to any operator. For twisted mass fermions, we use the
symmetry $S^u(x,y)=\gamma_5S^{d\dagger}(y,x)\gamma_5$ between the $u-$
and $d-$quark propagators. Therefore with a single inversion one can extract the vertex function for
 a {\em single} momentum.
The advantage of this approach is a high statistical
accuracy and the evaluation of the vertex 
for any operator including extended operators at
no significant additional computational cost. Since
 we are interested in a number
of operators with their associated renormalization constants 
we use the second approach. 
We fix to Landau gauge using a stochastic over-relaxation
algorithm~\cite{deForcrand:1989im}, converging to a gauge
transformation which minimizes the functional
\begin{equation}
   F = \sum_{x,\mu} {\rm Re}\ {\rm tr} \left[ U_\mu(x) + U^\dagger_\mu(x-\hat\mu)\right] \, .
\end{equation}
Questions related to the Gribov ambiguity will not be addressed in this work.
The propagator in momentum space, in the physical basis, is defined by
\begin{equation}\label{pprop}
   S^u(p) = \frac{a^8}{V}\sum_{x,y} e^{-ip(x-y)} \left\langle u(x) \overline u(y) \right\rangle\, , \qquad
   S^d(p) = \frac{a^8}{V}\sum_{x,y} e^{-ip(x-y)} \left\langle d(x) \overline d(y) \right\rangle \, .
\end{equation}
An amputated vertex function is given by
\begin{equation}
   \Gamma(p) = (S^u(p))^{-1}\, G(p)\, (S^d(p))^{-1} \, .
\end{equation}
and the corresponding renormalized quantities are
\begin{equation}
   S_R(p)      = Z_q S(p) \, , \qquad \qquad
   \Gamma_R(p) = Z_q^{-1} Z_\Op \Gamma(p) \, ,
\end{equation}
In the twisted basis at maximal twist, Eq.~(\ref{vfun}) takes the form
\begin{equation}\label{vfun_tm}
   G(p) = \frac{a^{12}}{4V}\sum_{x,y,z,z'} e^{-ip(x-y)} \left\langle(\eins+i\gamma_5) u(x) \overline u(z)(\eins+i\gamma_5) \J(z,z') (\eins-i\gamma_5) d(z')\
 \overline d(y)(\eins-i\gamma_5) \right\rangle \, .
\end{equation}
After integration over the fermion fields, and using $\s^u(x,z)=\gamma_5 {\s^d}^\dagger(z,x)\gamma_5$ this becomes
\begin{equation}
   G(p) = \frac{a^{12}}{4V} \sum_z \left\langle (\eins-i\gamma_5){\breve {\s^d}}^\dagger(z,p)(\eins-i\gamma_5) \J(z,z')
   (\eins-i\gamma_5)\breve \s^d(z,p)(\eins-i\gamma_5) \right\rangle^G \, ,
\end{equation}
where $\langle ... \rangle^G$ is the integration over gluon fields,
and $\breve \s(z,p) = \sum_y e^{ipy} \s(z,y)$ is the Fourier
transformed propagator on one of its argument on a particular gauge
background. It can be obtained by inversion using the Fourier source 
\begin{equation}
   b_\alpha^a(x) = e^{ipx} \delta_{\alpha \beta}\delta_{a b} \, ,
\end{equation}
for all Dirac $\alpha$ and color $a$ indices.
The propagators in the physical basis given in Eq.~(\ref{pprop}) can be obtained from
\begin{eqnarray}\label{pprop_tm}
   S^d(p) &=& \phantom{-}\frac{1}{4} \sum_z e^{-ipz} \langle (\eins-i\gamma_5)\breve \s^d(z,p)(\eins-i\gamma_5) \rangle^G \nonumber \\
   S^u(p) &=&           -\frac{1}{4} \sum_z e^{+ipz} \langle (\eins-i\gamma_5){\breve {\s^d}}^\dagger(z,p)(\eins-i\gamma_5) \rangle^G \, ,
\end{eqnarray}
%There are some subtleties related to the implementation of the boundary conditions in the HMC-code
%which we do not explain here.
which evidently only need 12 inversions despite the occurrence of both $u$ and $d$ quarks in the original
expression.

We evaluate Eq.~(\ref{vfun_tm}) and Eq.~(\ref{pprop_tm}) for each momentum
separately employing Fourier sources over a range of $a^2p^2$ for which
perturbative results can be trusted and finite $a$ corrections
are reasonably small. 
%Alternatively one could exploit
%translation invariance to shift the operator position in
%Eq.~(\ref{vfun_tm}) to position $z=0$ in each term. This would allow
%for an evaluation of the vertex function with all possible momenta at
%the cost of one set of inversions per configuration, but would lead to
%larger statistical errors. This second method has been carried out for
%local bilinears~\cite{Dimopoulos:2007fn} and one of the one-derivative
%operators~\cite{Zhaofeng} on the same configurations, leading to
%compatible results. 

%{\bf(more details...)}

\subsection{Renormalization Condition}
\label{subsectRC}
The renormalization constants are computed both perturbatively and
non-perturbatively in the RI$'$-MOM
scheme at different renormalization scales.% method of Ref.~\cite{Gockeler:1998ye}.
We translate them to the ${\overline{\rm MS}}$-scheme at (2~GeV)$^2$ using a conversion
factor computed in perturbation theory to ${\cal O}(g^6)$ as described in Section~\ref{sec5}.
The Z-factors are determined by imposing the following conditions:
\begin{eqnarray}
   Z_q = \frac{1}{12} {\rm Tr} \left[(S^L(p))^{-1}\, S^{(0)}(p)\right] \Bigr|_{p^2=\mu^2}  \\[2ex]
   Z_q^{-1}\,Z^{\mu\nu}_{\cal O}\,\frac{1}{12} {\rm Tr} \left[\Gamma^L_{\mu\nu}(p) \,\Gamma^{(0)-1}_{\mu\nu}(p)\right] \Bigr|_{p^2=\mu^2} &=& 1\, ,
\label{renormalization cond}
\end{eqnarray}
where $\mu$ is the renormalization scale, while $S_L$ and
$\Gamma_L$ correspond to the perturbative or non-perturbative
results. The trace is now taken over spin and color indices.
These conditions are imposed in the massless theory, i.e. at critical
mass and vanishing twisted mass.
At finite lattice spacing there are two choices
for $S^{(0)}$ and $\Gamma^{(0)}$ entering Eq.~(\ref{renormalization cond}). One can
take either the tree level or the continuum results for $S^{(0)}$ and
$\Gamma^{(0)}$, which differ by ${\cal O}(a^2)$-terms.
The continuum free propagator in terms of continuum
momentum is
\bea
\label{meth1}
    S^{(0)}(p) &=& \frac{-i \sum_\rho
     \gamma_\rho p_\rho}{p^2}\\ [1.5ex]
\Gamma^{(0)}_{\mu\nu}(p) &=& -i\, {\cal \tilde O}_{\{\mu} \,\, p_{\nu\}} \,.
\label{meth1a}
\eea
We refer to this choice as {\it method 1}.
A different choice is to define the free propagator
using the lattice momentum~\cite{Gockeler:1998ye,Zhaofeng}~:
\bea
    S^{(0)}(p) &=& \frac{-i
     \sum_\rho \gamma_\rho \sin(p_\rho)}{\sum_\rho\sin(p_\rho)^2}
\label{meth2a}
   \\ [1.5ex] 
\Gamma^{(0)}_{\mu\nu}(p) &=& -i\, {\cal \tilde O}_{\{\mu} \,\, \sin(p_{\nu\}})\, ,
\label{meth2}
\eea
which we will refer to as {\it method 2}  used in Ref.~\cite{Constantinou:2010gr}.
%where ${\cal \tilde O}_\mu$ is $\gamma_\mu$ ($\gamma_5\cdot\gamma_\mu$) for
%the vector (axial) operator.
\begin{figure}[h]
\centerline{\psfig{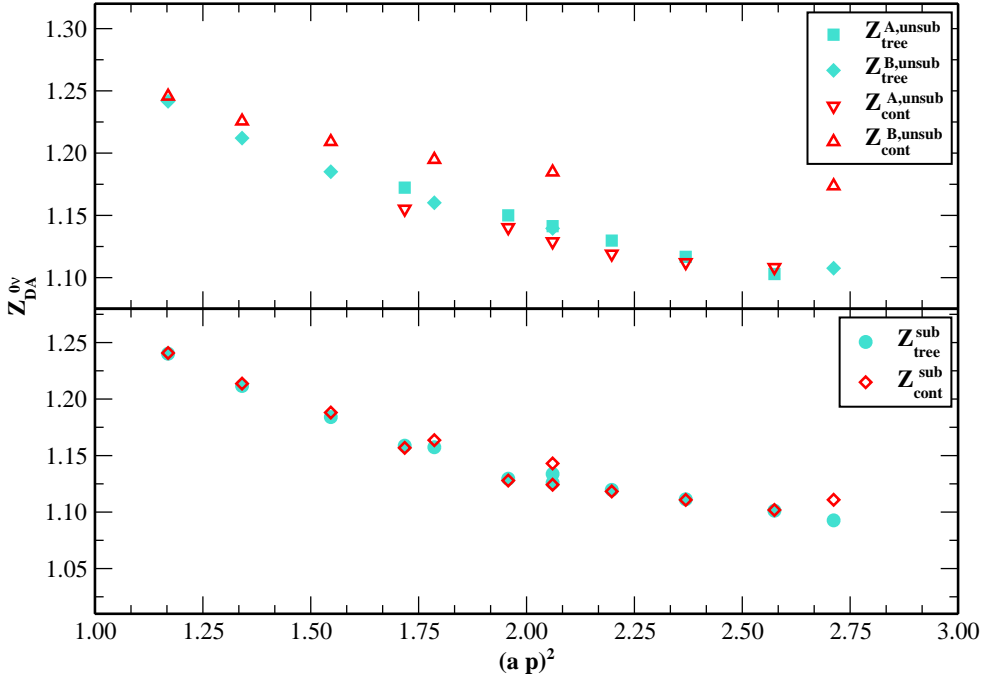}$\qquad\quad$}
\vspace{0.05cm}
\caption{$Z^{0\nu}_{\rm DA}$ for $\beta=3.9$ ($a^{-1}$=2.217~GeV) and
  $m_\pi=0.430$~GeV for method 1 (open symbols) and method 2 (filled
  symbols). The upper plot corresponds to non-perturbative
  results, where the index A, B represents the set of momenta with spatial
  components $2\,\pi/L\,(3,3,3)$ and $2\,\pi/L\,(2,2,2)$,
  respectively. The lower plot shows the non-perturbative results
  after subtracting the perturbative ${\cal O}(g^2\,a^2)$-terms, where the
  two methods give almost identical results. Moreover, in method 1,
  the jump between the two sets of momenta disappears.}
\label{fig3}
\end{figure}

\begin{figure}[h]
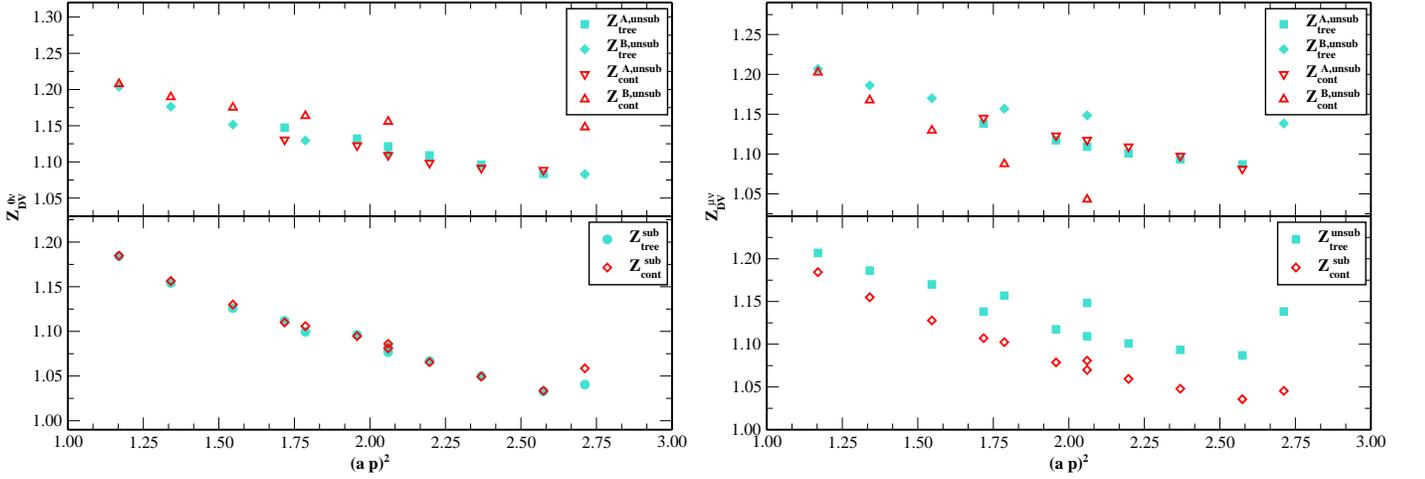

\centerline{\psfig{figure=Z_DV_0n_b3.9_0.0085_cont_tree.eps,height=6.3truecm,clip=}
      $\,\,$\psfig{figure=Z_DV_mn_b3.9_0.0085_cont_tree.eps,height=6.3truecm,clip=}}
\vspace{0.05cm}
\caption{In the left panel we show $Z^{0\nu}_{\rm DV}$  using the same notation as in Fig.~\ref{fig3}. In the right panel, upper graph we show  $Z^{\mu\nu}_{\rm DV}$ again using the notation of Fig.\ref{fig3} whereas in the lower graph we show a comparison between method 1 after subtracting
the perturbative ${\cal O}(a^2)$-terms (diamonds) and method 2 without any subtractions (filled squares).    }
\label{fig4}
\end{figure}

The choices for $S^{(0)}$ and $\Gamma^{(0)}_{\mu\nu}$ given in
Eqs.~(\ref{meth2a}) - (\ref{meth2}) are preferable, compared to
Eqs.~(\ref{meth1}) - (\ref{meth1a}), since only for method 2 we obtain
$Z_q = 1$, $Z_{\cal O} = 1$ when the gauge field is set to unity.
Similarly, in the perturbative computation only method 2 gives $Z_q =
1$, and $Z_{\cal O} = 1$ at tree-level. On the contrary, the Z-factors
obtained from method 1 have lattice artifacts even at tree-level. 
In this sense, method 2 is an improvement of method 1, and can
be thus considered superior.
Obviously, the renormalization constants using the two
methods differ only in their lattice artifacts, as can be seen by
Eq.~(\ref{RC2}). We find that, for the cases considered here, non-perturbative
results using method 2 lead to Z-factors with smaller lattice
effects. We demonstrate this by examining the following case: Let us consider
the momenta as given in Table~\ref{tab2}, which fall into two sets,
those with spatial components $2\,\pi\,(2,2,2)/L$ 
and those with $2\,\pi\,(3,3,3)/L$; there is only one non democratic
momentum, with spatial components $2\,\pi\,(3,3,2)/L$, but this
behaves similarly to the second set mentioned above. The two sets
of momenta do not fall on the same curve, a behavior that is due
to cut-off effects. This is clearly seen in
Fig.~\ref{fig3} where we present $Z^{0\nu}_{\rm DA}$ at $\beta=3.9$ and
$\mu_0=0.0085$ using the two methods (upper plot) (similarly for
$Z^{0\nu}_{\rm DV}$ in Fig.~\ref{fig4}). The statistical
errors in Fig.~\ref{fig3} and the rest of the graphs are  smaller than
the size of the symbols. As can be seen, 
 the two sets of momenta differ in particular when using method 1.
 However, it is important to note that,
after subtracting the ${\cal O}(a^2)$ perturbative contributions
 one
obtains values that are consistent between the two methods (see lower
plot). Moreover, in method 1 the jump observed between the two sets of momenta
disappears. 
For method 1 the subtraction of the perturbative ${\cal O}(a^2)$ 
terms refers to both contributions of $O(g^0\,a^2)$ (tree-level) and
$O(g^2\,a^2)$ (1-loop). Thus, {\it{upon subtraction}}  method 1 can be compared to method 2
because we remove all ${\cal O}(a^2)$ terms up to 1-loop, leading to
almost equivalent results as can be seen in the lower plot in Fig.~\ref{fig3}.
We would like to stress that the substraction is necessary 
and yields results superior to using the unsubtracted
method 2. This is demonstrated  in  Fig.~\ref{fig4} in the right
hand size plots where in the upper plot we show the unsubtracted methods 1 and 2,
while in the lower plot we show a comparison 
method 1 after subtraction of perturbative ${\cal O}(a^2)$-terms
 with the unsubtracted method 2. One can
observe that the jump between different set of momenta, appearing in
the unsubtracted method 2, almost disappears  in the subtracted method 1.
The same pattern appears in all Z-factors of the
one-derivative operators, as well as for $Z_q$. The latter has an
impact on all other renormalization constants discussed here. This
effect, as expected, becomes less pronounced at $\beta=4.05$ and
$4.20$, and disappears for small $a^2p^2$ as demonstrated in the next
section. The results presented in all Tables correspond to the
Z-factors obtained using method 2.

\section{Results}
\label{sec5}

\subsection{RI$'$-MOM condition}
We perform the calculation of renormalization constants for
 three values of the lattice
spacing corresponding to $\beta=3.9,\,4.05$ and $4.20$.
The lattice spacing as determined from the nucleon mass 
is 0.089~fm, 0.070~fm and 0.056~fm respectively. 
For $\beta=3.9$ we consider three different quark
masses, corresponding to $ m_\pi= 0.302$~GeV ($a\mu_0=0.004$), $m_\pi=0.376$~GeV
($a\mu_0=0.0064$) and $m_\pi=0.430$~GeV ($a\mu_0=0.0085$), in order
to explore the dependence of the Z-factors on the pion mass. At $\beta=4.05$
we consider two volumes, $24^3\times48$ and $32^3\times64$ in order to
check for finite volume effects. To extract the renormalization
constants reliably one needs to consider momenta in the range
$\Lambda_{QCD}<p<1/a$. We relax the upper bound to be $\sim 2/a$ to $3/a$, which
is justified by the linear dependence of our results on
$a^2$. Therefore, for each value of $\beta$ we consider momenta
spanning the range $1<a^2p^2<2.7$ for which perturbation theory is
trustworthy and lattice artifacts are still small enough. In
Table~\ref{tab1} we summarize the various parameters of the action,
that we used in our simulations.
\begin{table}[h]
\begin{center}
\begin{minipage}{15cm}
\begin{tabular}{ccccc}
\hline
\hline\\[-2.25ex]
$\beta$ & a (fm)& $a\mu_0$ & $m_\pi$~(GeV) & $L^3\times T$ \\ 
\hline
\hline\\[-2.25ex]
3.9  & 0.089  & 0.0040 &  0.3021(14)  &$24^3\times48$  \\
3.9  & 0.089  & 0.0064 &  0.37553(80) &$24^3\times48$  \\
3.9  & 0.089  & 0.0085 &  0.4302(11)  &$24^3\times48$  \\
4.05 & 0.070  & 0.006  &  0.4082(31)  &$24^3\times48$  \\
4.05 & 0.070  & 0.006  &  0.404(2)    &$32^3\times64$  \\
4.05 & 0.070  & 0.008  &  0.465(1)    &$32^3\times64$  \\
4.20 & 0.055  & 0.0065 &  0.476(2)    &$32^3\times64$  \\
\hline
\hline
\end{tabular}
\end{minipage}
\end{center}
\caption{Action parameters used in the simulations.} 
\label{tab1}
\end{table}

\noindent
In Table \ref{tab2} we present the statistical sample for the
parameters and momenta we used in the simulations. Using the number of
configurations shown in Table~\ref{tab2} leads to results with very
high statistical accuracy, easily below 0.5$\%$.

\begin{figure}
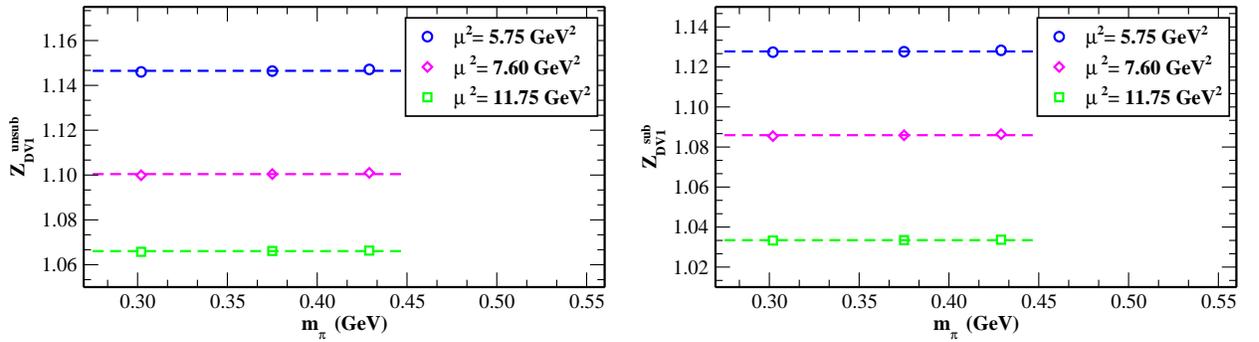

\centerline{\psfig{figure=Z_DV1_b3.9_mpi_dependence_unsub.eps,height=4.5truecm,clip=}
     $\quad$\psfig{figure=Z_DV1_b3.9_mpi_dependence.eps,height=4.5truecm,clip=}}
\vspace{0.1cm}
\caption{$Z_{\rm DV1}$ at $\beta=3.9$, as a function of the pion mass:
  $m_\pi=0.302$~GeV ($a\mu_0=0.004$), $m_\pi=0.375$~GeV
  ($a\mu_0=0.0064$) and $m_\pi=0.429$~GeV ($a\mu_0=0.0085$). The left
  plot regards the unsubtracted non-perturbative results and the right
one corresponds to the subtracted data. }
\label{fig5}
\end{figure}

{\small{
\begin{table}[h]
\begin{center}
\begin{minipage}{15cm}
\begin{tabular}{lr@{}lr@{}lr@{}lr@{}lr@{}lr@{}lr@{}l}
\hline
\hline\\[-2.25ex]
\multicolumn{1}{c}{$\phantom{A}^{\phantom{A^{\phantom{A}}}}_{\phantom{A_{\phantom{A}}}}$}&
\multicolumn{2}{c}{$\beta=3.9$} &
\multicolumn{2}{c}{$\beta=3.9$} &
\multicolumn{2}{c}{$\beta=3.9$} &
\multicolumn{2}{c}{$\beta=4.05$}&
\multicolumn{2}{c}{$\beta=4.05$}&
\multicolumn{2}{c}{$\beta=4.05$}&
\multicolumn{2}{c}{$\beta=4.20$} \\
\multicolumn{1}{c}{($n_t,n_x,n_y,n_z$)}&
\multicolumn{2}{c}{$24^3\times48$} &
\multicolumn{2}{c}{$24^3\times48$} &
\multicolumn{2}{c}{$24^3\times48$} &
\multicolumn{2}{c}{$24^3\times48$}&
\multicolumn{2}{c}{$32^3\times64$}&
\multicolumn{2}{c}{$32^3\times64$}&
\multicolumn{2}{c}{$32^3\times64$} \\
\multicolumn{1}{c}{$\phantom{A}^{\phantom{A^{\phantom{A}}}}_{\phantom{A_{\phantom{A}}}}$}&
\multicolumn{2}{c}{$\mu_0=0.004\,\,$} &
\multicolumn{2}{c}{$\mu_0=0.0064\,\,$} &
\multicolumn{2}{c}{$\mu_0=0.0085\,\,$} &
\multicolumn{2}{c}{$\mu_0=0.006\,\,$}&
\multicolumn{2}{c}{$\mu_0=0.006\,\,$}&
\multicolumn{2}{c}{$\mu_0=0.008\,\,$}&
\multicolumn{2}{c}{$\mu_0=0.0065$} \\
\hline
\hline\\[-2.25ex]
(4,2,2,2)  &$\,\,\quad$&100  &$\,\,\quad$&50     &$\,\,\quad$&80   &$\,\,\quad$&---    &$\,\,\quad$&50   &$\,\,\quad$&50    &$\,\,\quad$&15$^{\phantom{A^{\phantom{A}}}}$\\
(5,2,2,2)  &$\,\,\quad$&100  &$\,\,\quad$&60     &$\,\,\quad$&60   &$\,\,\quad$&---    &$\,\,\quad$&---  &$\,\,\quad$&33    &$\,\,\quad$&15                       \\
(6,2,2,2)  &$\,\,\quad$&100  &$\,\,\quad$&50     &$\,\,\quad$&50   &$\,\,\quad$&---    &$\,\,\quad$&---  &$\,\,\quad$&50    &$\,\,\quad$&15                       \\
(3,3,3,2)  &$\,\,\quad$&---  &$\,\,\quad$&---    &$\,\,\quad$&27   &$\,\,\quad$&---    &$\,\,\quad$&---  &$\,\,\quad$&15    &$\,\,\quad$&15                       \\
(7,2,2,2)  &$\,\,\quad$&---  &$\,\,\quad$&---    &$\,\,\quad$&20   &$\,\,\quad$&---    &$\,\,\quad$&---  &$\,\,\quad$&15    &$\,\,\quad$&15                       \\
(2,3,3,3)  &$\,\,\quad$&---  &$\,\,\quad$&---    &$\,\,\quad$&20   &$\,\,\quad$&---    &$\,\,\quad$&---  &$\,\,\quad$&15    &$\,\,\quad$&15                       \\
(8,2,2,2)  &$\,\,\quad$&---  &$\,\,\quad$&---    &$\,\,\quad$&20   &$\,\,\quad$&---    &$\,\,\quad$&---  &$\,\,\quad$&15    &$\,\,\quad$&15                       \\
(3,3,3,3)  &$\,\,\quad$&100  &$\,\,\quad$&50     &$\,\,\quad$&80   &$\,\,\quad$&15     &$\,\,\quad$&---  &$\,\,\quad$&50    &$\,\,\quad$&15                       \\
(4,4,4,4)  &$\,\,\quad$&---  &$\,\,\quad$&---    &$\,\,\quad$&---  &$\,\,\quad$&---    &$\,\,\quad$&15   &$\,\,\quad$&---   &$\,\,\quad$&---                      \\
(4,3,3,3)  &$\,\,\quad$&100  &$\,\,\quad$&60     &$\,\,\quad$&60   &$\,\,\quad$&---    &$\,\,\quad$&---  &$\,\,\quad$&50    &$\,\,\quad$&15                       \\
(5,3,3,3)  &$\,\,\quad$&100  &$\,\,\quad$&60     &$\,\,\quad$&60   &$\,\,\quad$&---    &$\,\,\quad$&---  &$\,\,\quad$&50    &$\,\,\quad$&15                       \\
(6,3,3,3)  &$\,\,\quad$&---  &$\,\,\quad$&---    &$\,\,\quad$&15   &$\,\,\quad$&---    &$\,\,\quad$&---  &$\,\,\quad$&15    &$\,\,\quad$&15                       \\
(10,2,2,2) &$\,\,\quad$&---  &$\,\,\quad$&---    &$\,\,\quad$&15   &$\,\,\quad$&---    &$\,\,\quad$&---  &$\,\,\quad$&15    &$\,\,\quad$&15                       \\
(8,3,3,3)  &$\,\,\quad$&---  &$\,\,\quad$&---    &$\,\,\quad$&---  &$\,\,\quad$&---    &$\,\,\quad$&---  &$\,\,\quad$&15    &$\,\,\quad$&15                       \\
(9,3,3,3)  &$\,\,\quad$&---  &$\,\,\quad$&---    &$\,\,\quad$&---  &$\,\,\quad$&---    &$\,\,\quad$&---  &$\,\,\quad$&15    &$\,\,\quad$&15                       \\
(10,3,3,3) &$\,\,\quad$&---  &$\,\,\quad$&---    &$\,\,\quad$&---  &$\,\,\quad$&---    &$\,\,\quad$&---  &$\,\,\quad$&15    &$\,\,\quad$&15                       \\
(13,2,2,2) &$\,\,\quad$&---  &$\,\,\quad$&---    &$\,\,\quad$&---  &$\,\,\quad$&---    &$\,\,\quad$&---  &$\,\,\quad$&15    &$\,\,\quad$&15                       \\
(11,3,3,3) &$\,\,\quad$&---  &$\,\,\quad$&---    &$\,\,\quad$&---  &$\,\,\quad$&---    &$\,\,\quad$&---  &$\,\,\quad$&15    &$\,\,\quad$&15                       \\
(14,2,2,2) &$\,\,\quad$&---  &$\,\,\quad$&---    &$\,\,\quad$&---  &$\,\,\quad$&---    &$\,\,\quad$&---  &$\,\,\quad$&15    &$\,\,\quad$&15                        \\
(7,4,4,4)  &$\,\,\quad$&---  &$\,\,\quad$&---    &$\,\,\quad$&15   &$\,\,\quad$&---    &$\,\,\quad$&---  &$\,\,\quad$&---   &$\,\,\quad$&---                       \\
(8,4,4,4)  &$\,\,\quad$&---  &$\,\,\quad$&---    &$\,\,\quad$&15   &$\,\,\quad$&---    &$\,\,\quad$&---  &$\,\,\quad$&15    &$\,\,\quad$&---                       \\
(9,4,4,4)  &$\,\,\quad$&---  &$\,\,\quad$&---    &$\,\,\quad$&15   &$\,\,\quad$&---    &$\,\,\quad$&---  &$\,\,\quad$&15    &$\,\,\quad$&---                       \\
(10,4,4,4)  &$\,\,\quad$&---  &$\,\,\quad$&---    &$\,\,\quad$&--- &$\,\,\quad$&---    &$\,\,\quad$&---  &$\,\,\quad$&15    &$\,\,\quad$&---                       \\
(11,2,2,2) &$\,\,\quad$&---  &$\,\,\quad$&---    &$\,\,\quad$&15   &$\,\,\quad$&---    &$\,\,\quad$&---  &$\,\,\quad$&---   &$\,\,\quad$&---                       \\
(12,2,2,2) &$\,\,\quad$&---  &$\,\,\quad$&---    &$\,\,\quad$&15   &$\,\,\quad$&---    &$\,\,\quad$&---  &$\,\,\quad$&---   &$\,\,\quad$&---                       \\
(12,3,3,3) &$\,\,\quad$&---  &$\,\,\quad$&---    &$\,\,\quad$&15   &$\,\,\quad$&---    &$\,\,\quad$&---  &$\,\,\quad$&---   &$\,\,\quad$&---                       \\
(13,3,3,3) &$\,\,\quad$&---  &$\,\,\quad$&---    &$\,\,\quad$&---  &$\,\,\quad$&---    &$\,\,\quad$&---  &$\,\,\quad$&15    &$\,\,\quad$&---                       \\
(14,3,3,3) &$\,\,\quad$&---  &$\,\,\quad$&---    &$\,\,\quad$&---  &$\,\,\quad$&---    &$\,\,\quad$&---  &$\,\,\quad$&15    &$\,\,\quad$&---$_{\phantom{A_{\phantom{A}}}}$\\
\hline
\end{tabular}
\end{minipage}
\end{center}
\caption{Statistical sample at $\beta=3.9,\,4.05,\,4.20$ for various momenta.}
\label{tab2}
\end{table}
}}
\noindent
The results for the subtracted Z-factors (method 2) at $\beta=3.9$ are tabulated in
Table III for the highest and lowest twisted mass parameter used (for
the lowest mass we have obtained the Z-factors only for 6 momenta).
Comparison between the $Z-$factors for  two different  masses 
shows that any dependence on the pion mass is within the small statistical errors.
 This negligible dependence is not a result of the
${\cal O}(a^2)$ subtraction, as demonstrated in Fig.~\ref{fig5}. The
left plot illustrates the pion mass dependence of the unsubtracted
$Z_{\rm DV1}$ for three renormalization scales ranging from
5.75~GeV$^2$ to 11.75~GeV$^2$, while the subtracted $Z_{\rm DV1}$ is
shown in the right plot. The same behavior is observed for all
renormalization constants considered here. The subtracted Z-factors (method 2) for
$\beta=4.05$ and $\beta=4.20$ are presented in Tables \ref{tab4}-\ref{tab5},
respectively. In order to see possible volume effects we compute the
renormalization constants at $\beta=4.05,\,\mu_0=0.006$, for two lattices with
different size, namely for $24^3\times48$ and for $32^3\times
64$. For this comparison we used momenta that correspond to the same
renormalization scale: For the small lattice we use
$2\pi(3/48,3/24,3/24,3/24)$, in lattice units, whereas for the larger one we employ
$2\pi(4/64,4/32,4/32,4/32)$. The volume effects appear to be $\sim$
0.1$\%$, as can be seen from Table \ref{tab6}.

{\small{
\begin{table}[h]
\label{tab3}
\hspace{-0.5cm}
\begin{center}
\hspace{-0.75cm}
$\hspace{2cm} \mu_0=0.004 \hspace{5.7cm} \mu_0=0.0085$\\
\vspace{0.35cm}
\hspace{-0.5cm}
\begin{minipage}{1.85cm}
\hspace{-0.5cm}
\begin{tabular}{c}
\hline
\hline
\multicolumn{1}{c}{$(n_t,n_x,n_y,n_z)$}\\
\hline
\hline
(4,2,2,2)   \\
(5,2,2,2)   \\
(6,2,2,2)   \\
(3,3,3,2)   \\
(7,2,2,2)   \\
(2,3,3,3)   \\
(8,2,2,2)   \\
(3,3,3,3)   \\
(4,3,3,3)   \\
(5,3,3,3)   \\
(6,3,3,3)   \\
(10,2,2,2)  \\
\hline
\hline
\end{tabular}
\end{minipage}
\begin{minipage}{6.75cm}
\begin{tabular}{lr@{}lr@{}lr@{}l}
\hline
\hline
\multicolumn{1}{c}{$\,\,Z_{\rm DV1}\,\,$} &
\multicolumn{2}{c}{$\,\,Z_{\rm DV2}\,\,$} &
\multicolumn{2}{c}{$\,\,Z_{\rm DA1}\,\,$} &
\multicolumn{2}{c}{$\,\,Z_{\rm DA2}\,\,$}  \\
\hline
\hline
1.1274(1)   &$\,1$.&1836(4)   &$\,1$.&2044(2)   &$\,1$.&2387(4)       \\
1.1058(1)   &$\,1$.&1548(4)   &$\,1$.&1792(2)   &$\,1$.&2094(4)       \\
1.0854(1)   &$\,1$.&1283(3)   &$\,1$.&1567(2)   &$\,1$.&1820(3)       \\
---   &---&   &---&   &---&       \\
---   &---&   &---&   &---&       \\
---   &---&   &---&   &---&       \\
---   &---&   &---&   &---&       \\
---   &---&   &---&   &---&       \\
1.04985(7)   &$\,1$.&0750(3)   &$\,1$.&1100(1)    &$\,1$.&1315(3)       \\
1.04152(6)   &$\,1$.&0620(2)   &$\,1$.&09035(9)   &$\,1$.&1223(2)       \\
1.03327(5)   &$\,1$.&0482(2)   &$\,1$.&07332(8)   &$\,1$.&1120(2)       \\
---   &---&   &---&   &---&       \\
\hline
\hline
\end{tabular}
\end{minipage}
%\end{center}
%
%\begin{center}
\begin{minipage}{6.75cm}
\begin{tabular}{lr@{}lr@{}lr@{}l}
\hline
\hline
\multicolumn{1}{c}{$\,\,Z_{\rm DV1}\,\,$} &
\multicolumn{2}{c}{$\,\,Z_{\rm DV2}\,\,$} &
\multicolumn{2}{c}{$\,\,Z_{\rm DA1}\,\,$} &
\multicolumn{2}{c}{$\,\,Z_{\rm DA2}\,\,$}  \\
\hline
\hline
1.1283(2)    &$\,1$.&1846(5)   &$\,1$.&2051(2)    &$\,1$.&2395(5)       \\
1.1067(2)    &$\,1$.&1558(5)   &$\,1$.&1800(2)    &$\,1$.&2102(5)       \\
1.0864(1)    &$\,1$.&1291(4)   &$\,1$.&1576(2)    &$\,1$.&1829(4)       \\
1.0740(1)    &$\,1$.&1088(4)   &$\,1$.&1418(2)    &$\,1$.&1626(4)       \\
1.06613(8)   &$\,1$.&1042(4)   &$\,1$.&1363(2)    &$\,1$.&1568(4)       \\
1.0587(1)    &$\,1$.&0869(4)   &$\,1$.&1331(2)    &$\,1$.&1391(4)       \\
1.04684(6)   &$\,1$.&0830(3)   &$\,1$.&1176(1)    &$\,1$.&1341(3)       \\
1.05045(6)   &$\,1$.&0756(2)   &$\,1$.&11066(9)   &$\,1$.&1321(2)       \\
1.04204(6)   &$\,1$.&0625(2)   &$\,1$.&09097(9)   &$\,1$.&1229(2)       \\
1.03367(6)   &$\,1$.&0487(2)   &$\,1$.&07377(9)   &$\,1$.&1124(2)       \\
1.02513(5)   &$\,1$.&0346(2)   &$\,1$.&05817(6)   &$\,1$.&1009(1)       \\
1.00804(9)   &$\,1$.&0482(2)   &$\,1$.&08561(9)   &$\,1$.&0945(2)       \\
\hline
\hline
\end{tabular}
\end{minipage}
\end{center}
{\normalsize{\caption{The renormalization constants at $\beta=3.9$ with
      $\mu_0=0.004,\,0.0085$ for lattice size:
      $24^3\times48$.}}}
\end{table}
}}

\vspace{1cm}

\begin{table}[h]
\begin{center}
\begin{minipage}{15cm}
\begin{tabular}{lr@{}lr@{}lr@{}lr@{}l}
\hline
\hline\\[-2.25ex]
\multicolumn{1}{c}{$(n_t,n_x,n_y,n_z)$}&
\multicolumn{2}{c}{$\,\,Z_{\rm DV1}\,\,$} &
\multicolumn{2}{c}{$\,\,Z_{\rm DV2}\,\,$} &
\multicolumn{2}{c}{$\,\,Z_{\rm DA1}\,\,$} &
\multicolumn{2}{c}{$\,\,Z_{\rm DA2}\,\,$}  \\
\hline
\hline\\[-2.25ex]
$\quad$(4,2,2,2) $\,$ &$\,\,\,1$.&1960(1)    &$\,\,\,1$.&2644(3)   &$\,\,\,1$.&2749(1)    &$\,\,\,1$.&3126(3)       \\
$\quad$(5,2,2,2) $\,$ &$\,\,\,1$.&1718(2)    &$\,\,\,1$.&2324(5)   &$\,\,\,1$.&2483(2)    &$\,\,\,1$.&2794(5)       \\
$\quad$(6,2,2,2) $\,$ &$\,\,\,1$.&1491(1)    &$\,\,\,1$.&2016(2)   &$\,\,\,1$.&2244(1)    &$\,\,\,1$.&2475(2)       \\
$\quad$(3,3,3,2) $\,$ &$\,\,\,1$.&1336(1)    &$\,\,\,1$.&1805(3)   &$\,\,\,1$.&2069(1)    &$\,\,\,1$.&2260(3)       \\
$\quad$(7,2,2,2) $\,$ &$\,\,\,1$.&1280(2)    &$\,\,\,1$.&1745(2)   &$\,\,\,1$.&2025(2)    &$\,\,\,1$.&2200(2)       \\
$\quad$(2,3,3,3) $\,$ &$\,\,\,1$.&1188(1)    &$\,\,\,1$.&1555(3)   &$\,\,\,1$.&1948(1)    &$\,\,\,1$.&1998(3)       \\
$\quad$(8,2,2,2) $\,$ &$\,\,\,1$.&1086(1)    &$\,\,\,1$.&1493(2)   &$\,\,\,1$.&1826(2)    &$\,\,\,1$.&1931(2)       \\
$\quad$(3,3,3,3) $\,$ &$\,\,\,1$.&10592(6)   &$\,\,\,1$.&1458(1)   &$\,\,\,1$.&17497(7)   &$\,\,\,1$.&1914(2)       \\
$\quad$(4,3,3,3) $\,$ &$\,\,\,1$.&09370(5)   &$\,\,\,1$.&1339(1)   &$\,\,\,1$.&15739(6)   &$\,\,\,1$.&1807(1)       \\
$\quad$(5,3,3,3) $\,$ &$\,\,\,1$.&08232(5)   &$\,\,\,1$.&1206(1)   &$\,\,\,1$.&14186(6)   &$\,\,\,1$.&1683(1)       \\
$\quad$(6,3,3,3) $\,$ &$\,\,\,1$.&07144(9)   &$\,\,\,1$.&1068(2)   &$\,\,\,1$.&1278(1)    &$\,\,\,1$.&1551(2)       \\
$\quad$(10,2,2,2)     &$\,\,\,1$.&0724(1)    &$\,\,\,1$.&1090(2)   &$\,\,\,1$.&1477(1)    &$\,\,\,1$.&1505(2)       \\
\hline
\hline
\end{tabular}
\end{minipage}
\end{center}
\caption{Renormalization constants at $\beta=4.05, a\mu_0=0.008$ for lattice 
size $32^3\times64$.}
\label{tab4}
\end{table}

{\small{
\begin{table}[h]
\begin{center}
\begin{minipage}{15cm}
\begin{tabular}{lr@{}lr@{}lr@{}lr@{}l}
\hline
\hline\\[-2.25ex]
\multicolumn{1}{c}{$(n_t,n_x,n_y,n_z)$}&
\multicolumn{2}{c}{$\,\,Z_{\rm DV1}\,\,$} &
\multicolumn{2}{c}{$\,\,Z_{\rm DV2}\,\,$} &
\multicolumn{2}{c}{$\,\,Z_{\rm DA1}\,\,$} &
\multicolumn{2}{c}{$\,\,Z_{\rm DA2}\,\,$}  \\
\hline
\hline\\[-2.25ex]
$\quad$(4,2,2,2) $\,$ &$\,\,\,1$.&1585(4)   &$\,\,\,1$.&215(1)    &$\,\,\,1$.&2266(5)   &$\,\,\,1$.&257(1)        \\
$\quad$(5,2,2,2) $\,$ &$\,\,\,1$.&1387(4)   &$\,\,\,1$.&189(1)    &$\,\,\,1$.&2052(5)   &$\,\,\,1$.&230(1)        \\
$\quad$(6,2,2,2) $\,$ &$\,\,\,1$.&1203(3)   &$\,\,\,1$.&1642(9)   &$\,\,\,1$.&1853(4)   &$\,\,\,1$.&2040(9)       \\
$\quad$(3,3,3,2) $\,$ &$\,\,\,1$.&1069(2)   &$\,\,\,1$.&1459(9)   &$\,\,\,1$.&1702(3)   &$\,\,\,1$.&1855(9)       \\
$\quad$(7,2,2,2) $\,$ &$\,\,\,1$.&1028(2)   &$\,\,\,1$.&1413(8)   &$\,\,\,1$.&1668(3)   &$\,\,\,1$.&1804(8)       \\
$\quad$(2,3,3,3) $\,$ &$\,\,\,1$.&0943(2)   &$\,\,\,1$.&1257(8)   &$\,\,\,1$.&1599(3)   &$\,\,\,1$.&1643(8)       \\
$\quad$(8,2,2,2) $\,$ &$\,\,\,1$.&0853(1)   &$\,\,\,1$.&1197(4)   &$\,\,\,1$.&1488(2)   &$\,\,\,1$.&1579(4)       \\
$\quad$(3,3,3,3) $\,$ &$\,\,\,1$.&0841(1)   &$\,\,\,1$.&1177(8)   &$\,\,\,1$.&1438(3)   &$\,\,\,1$.&1577(8)       \\
$\quad$(4,3,3,3) $\,$ &$\,\,\,1$.&0743(2)   &$\,\,\,1$.&1079(7)   &$\,\,\,1$.&1293(2)   &$\,\,\,1$.&1491(7)       \\
$\quad$(5,3,3,3) $\,$ &$\,\,\,1$.&0651(2)   &$\,\,\,1$.&0968(6)   &$\,\,\,1$.&1163(2)   &$\,\,\,1$.&1390(6)       \\
$\quad$(6,3,3,3) $\,$ &$\,\,\,1$.&0511(2)   &$\,\,\,1$.&0812(5)   &$\,\,\,1$.&0983(3)   &$\,\,\,1$.&1274(5)       \\
$\quad$(10,2,2,2)     &$\,\,\,1$.&0528(1)   &$\,\,\,1$.&0856(3)   &$\,\,\,1$.&1189(1)   &$\,\,\,1$.&1223(3)       \\
\hline
\hline
\end{tabular}
\end{minipage}
\end{center}
\caption{Renormalization constants at $\beta=4.20, \mu_0=0.0065$ for lattice size: $32^3\times64$. }
\label{tab5}
\end{table}

}}
\begin{figure}
\centerline{\psfig{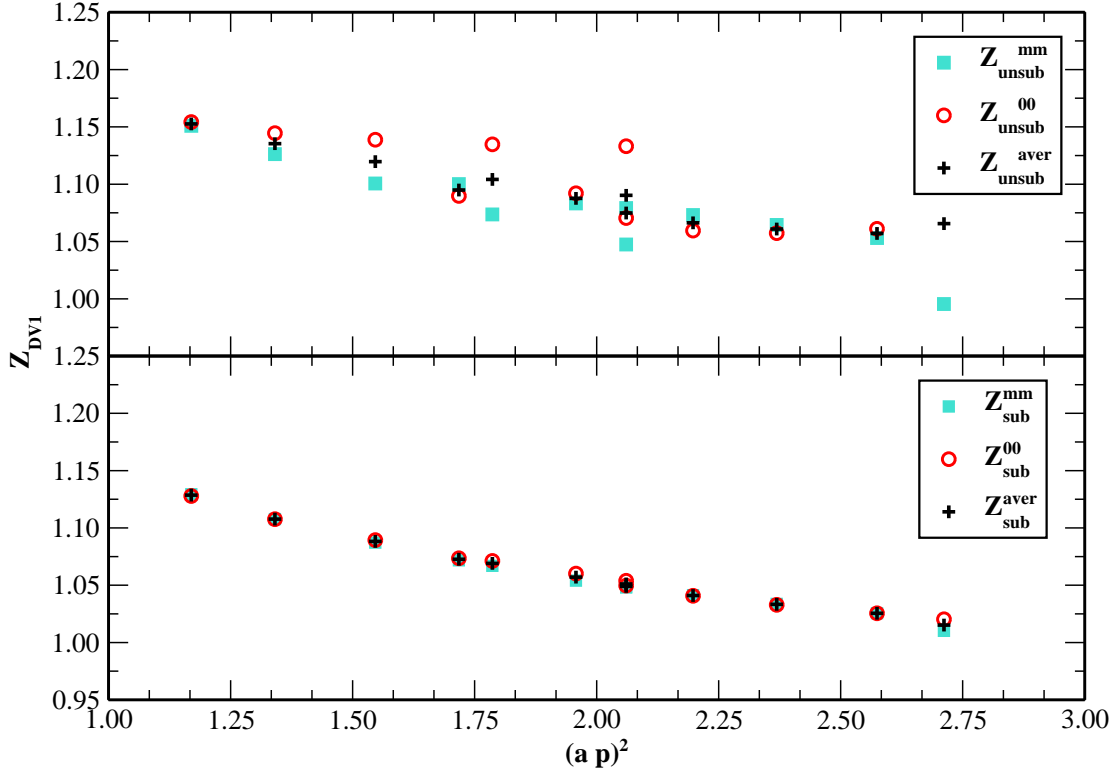}$\qquad\quad\quad\,$}
\vspace{0.1cm}
\caption{$Z^{\mu\mu}_{\rm DV}$ (squares), $Z^{00}_{\rm DV}$ (circles),
  $Z^{\rm aver}_{\rm DV}$ (crosses), for $\beta=3.9$
  ($a^{-1}$=2.217~GeV), $m_\pi=0.430$~GeV using method 1. The upper
  plot corresponds to the purely non-perturbative results, while the
  lower plot shows the non-perturbative results after subtracting the
  perturbative terms of ${\cal O}(a^2)$.}
\label{fig6}
\end{figure}
Given the small statistical errors one may carefully examine the
systematic errors. As already noted a systematic effect comes from the
choice of $S^{(0)}$ and $\Gamma^{(0)}$. To give an example, at
$\beta=3.9,\ \mu_0=0.004,\ \mu^2 \approx 5.75$~GeV$^2$ method~1 leads
to $Z_q=0.76606(7)$ while method~2 gives $Z_q=0.80514(7)$, before any
subtraction of ${\cal O}(a^2)$ is carried out. This systematic effect
is removed after perturbative subtraction is applied. Another, much
smaller, systematic effect comes from the asymmetry of our lattices
both because they are larger in their time extent and because of the
antiperiodic boundary conditions in the time direction. For instance,
using the same $\beta$ and $\mu_0$ as in the previous example, method
1 at $\mu^2\approx $7.6~GeV$^2$, in the temporal direction of the
current gives $Z_{\rm DV1}=1.1387(2)$ while the average from the three
spatial directions leads to $Z_{\rm DV1}=1.1006(2)$. This effect can
be seen in Fig. \ref{fig6} where we plot separately the
renormalization constant $Z_{\rm DV1}$ determined from the temporal
indices, the spatial indices and the average of those two. In the same
figure we also show that upon subtraction this systematic effect
disappears (lower plot). For Tables~III - \ref{tab5} we use for
$Z_{\rm DV1}$ the average of $Z_{\rm DV}^{00},\,Z_{\rm DV}^{\nu\nu}$
with $\nu=1,2,3$, while for $Z_{\rm DV2}$ the average of $Z_{\rm
  DV}^{0\nu},\,Z_{\rm DV}^{\nu\rho}$ with $\nu\neq\rho=1,2,3$. We
apply the same procedure for the twist-2 axial operator.

\begin{figure}
\centerline{\psfig{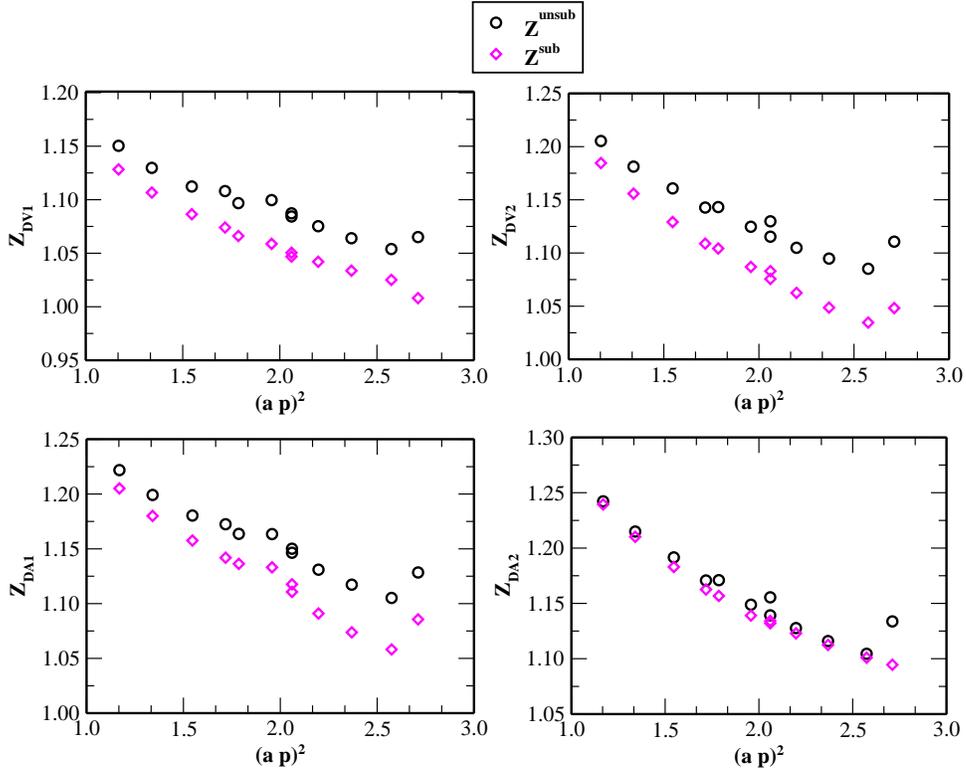}$\qquad\quad$}
\vspace{0.1cm}
\caption{Renormalization scale dependence for the Z-factors at $\beta=3.9$ and $m_\pi=0.430$~GeV}
\label{fig8}
\end{figure}

\begin{figure}
\centerline{\psfig{figure=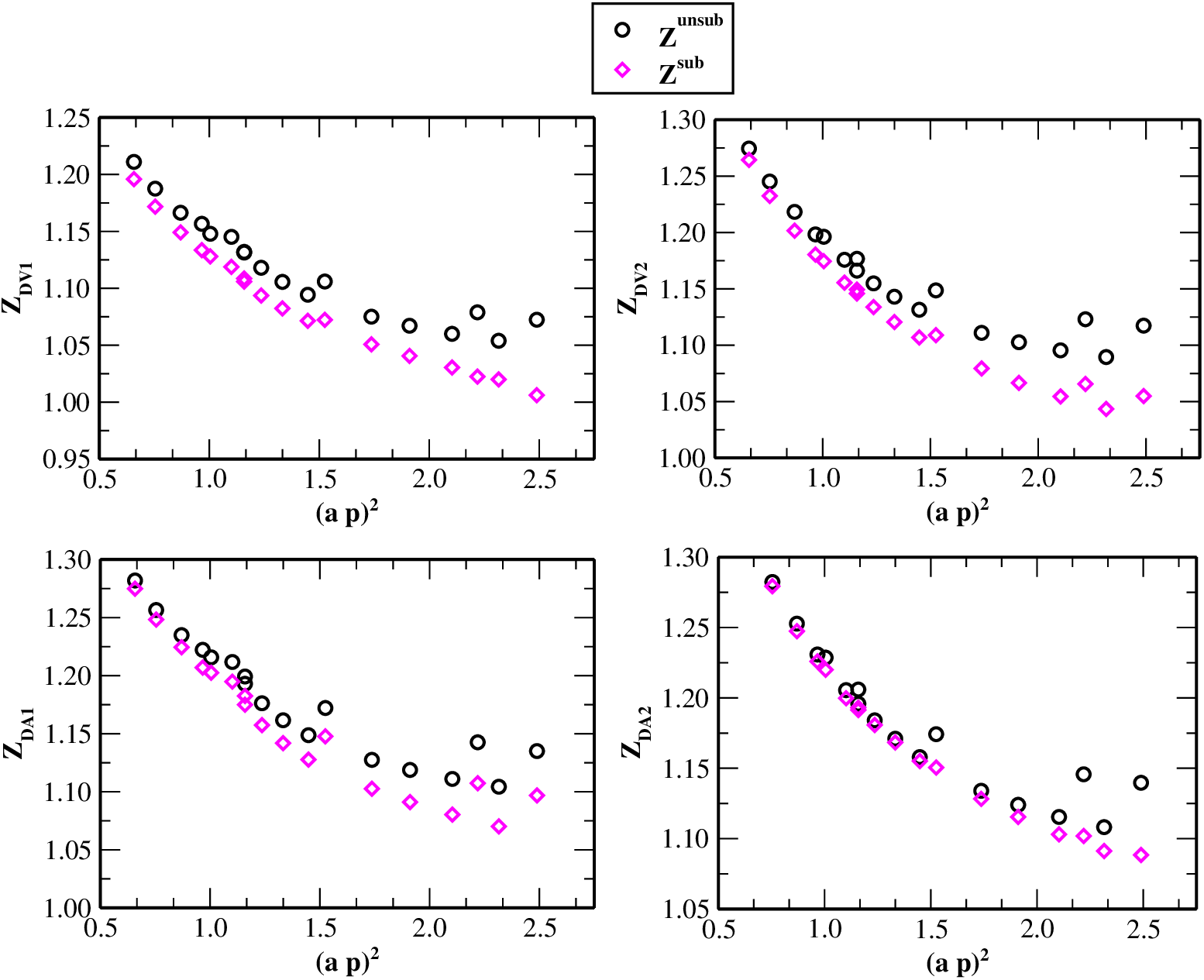,height=10.25truecm,clip=}$\qquad\quad$}
\vspace{0.1cm}
\caption{Renormalization scale dependence for the Z-factors at $\beta=4.05$ and $m_\pi=0.465$~GeV}
\label{fig13}
\end{figure}

\begin{figure}
\centerline{\psfig{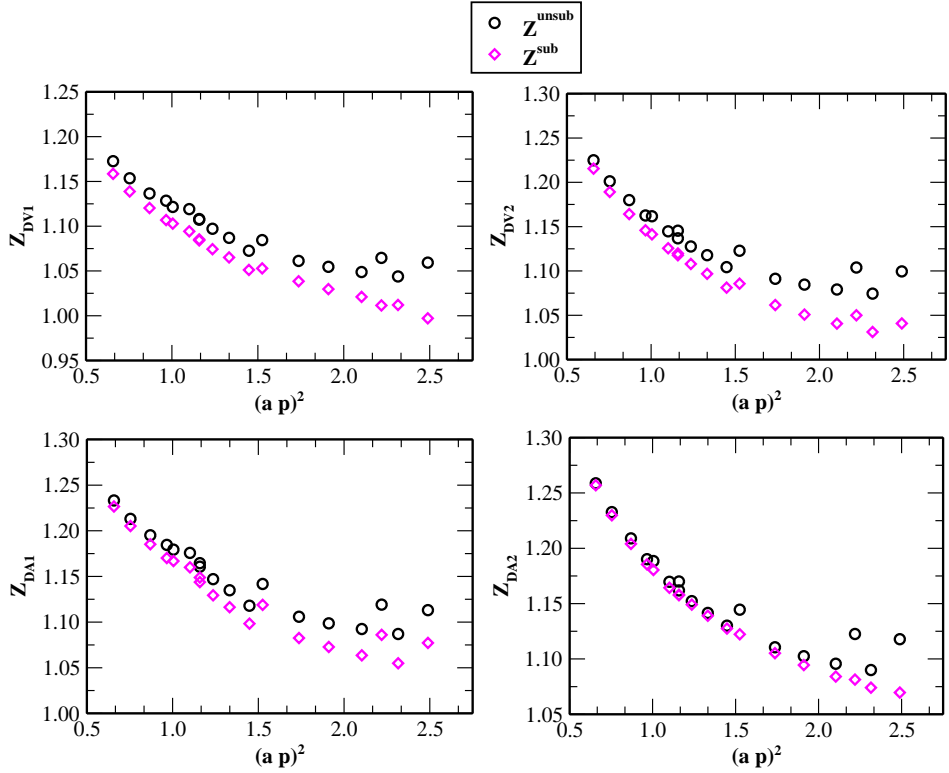}$\qquad\quad$}
\vspace{0.1cm}
\caption{Renormalization scale dependence for the Z-factors at $\beta=4.20$ and $m_\pi=0.476$~GeV}
\label{fig14}
\end{figure}

Chiral extrapolations are necessary to obtain the renormalization
factors in the chiral limit. As already pointed out the dependence on
the pion mass is insignificant. Allowing a slope and performing a
linear extrapolation to the data shown in Fig.~\ref{fig5} yields a
slope consistent with zero. This behavior is also observed at the
other $\beta$-values and therefore the renormalization constants are
computed at one quark mass, given in the Tables~III-\ref{tab5}. 
Figures \ref{fig8}, \ref{fig13}, \ref{fig14} demonstrate the effect of
subtraction, for all three $\beta$ values, as a function of the
renormalization scale (in lattice units). For all cases  we observe a significant correction upon subtraction; the 
lattice artifacts for $Z_{\rm DA2}$ turn out to be very small for
most values of the momentum. 
In addition, the lattice artifacts decrease by employing higher values
for $\beta$ (finer lattice), as expected. 

\begin{table}[h]
\begin{center}
\begin{minipage}{15cm}
\begin{tabular}{lr@{}lr@{}lr@{}lr@{}l}
\hline
\hline\\[-2.25ex]
\multicolumn{1}{c}{lattice}&
\multicolumn{2}{c}{$\,\,Z_{\rm DV1}\,\,$} &
\multicolumn{2}{c}{$\,\,Z_{\rm DV2}\,\,$} &
\multicolumn{2}{c}{$\,\,Z_{\rm DA1}\,\,$} &
\multicolumn{2}{c}{$\,\,Z_{\rm DA2}\,\,$}  \\
\hline
\hline\\[-2.25ex]
$24^3$x$48$  &$\,\,\,1$.&0700(2)    &$\,\,\,1$.&0923(2)   &$\,\,\,1$.&1190(2)    &$\,\,\,1$.&1117(2)       \\
$32^3$x$64$  &$\,\,\,1$.&07123(6)   &$\,\,\,1$.&0928(2)   &$\,\,\,1$.&12037(7)   &$\,\,\,1$.&1122(2)       \\
\hline
\hline
\end{tabular}
\end{minipage}
\end{center}
\caption{Renormalization constants at $\beta=4.05, \mu_0=0.008$
 using
  method 2 and two lattice sizes: $32^3\times64$ for (4,4,4,4) and $24^3$x$48$
  for the rest of the momenta.}
\label{tab6}
\end{table}

\begin{figure}[h]
\centerline{\psfig{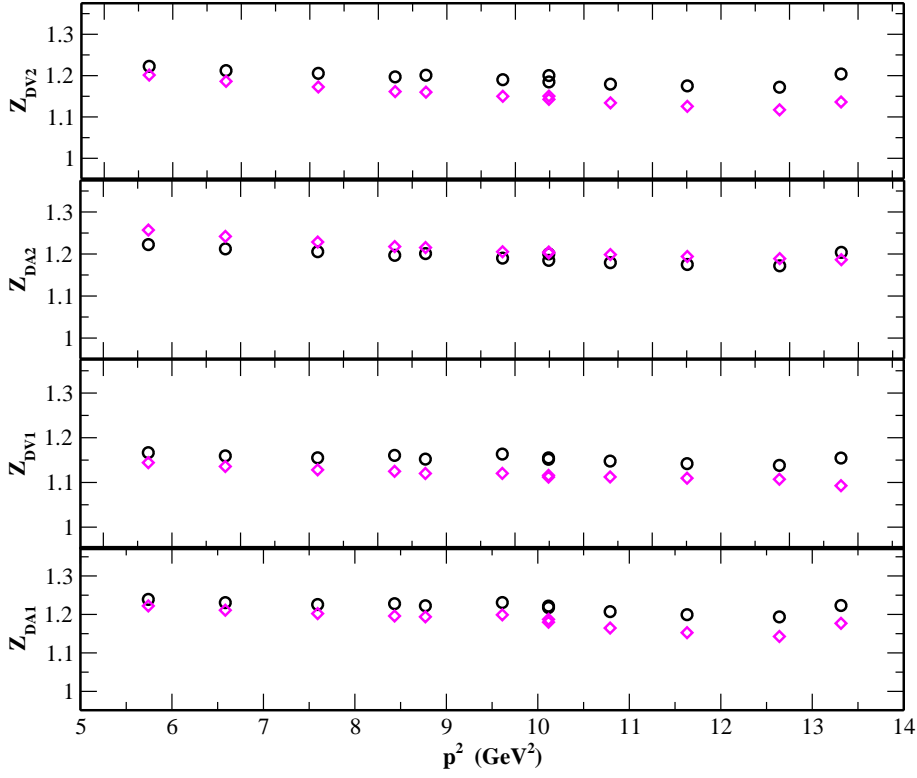}$\qquad\quad$}
\vspace{0.1cm}
   \caption{ Renormalization factors in the RI$'$-MOM scheme
   at renormalization scale $1/a$, for $\beta=3.9,\,\mu_0=0.0085$. The black circles correspond 
   to the unsubtracted results, while the magenta diamonds to the
   results with perturbatively subtracted one loop $O(a^2)$
   artifacts.}
\label{fig9}
\end{figure}

\begin{figure}[h]
\centerline{\psfig{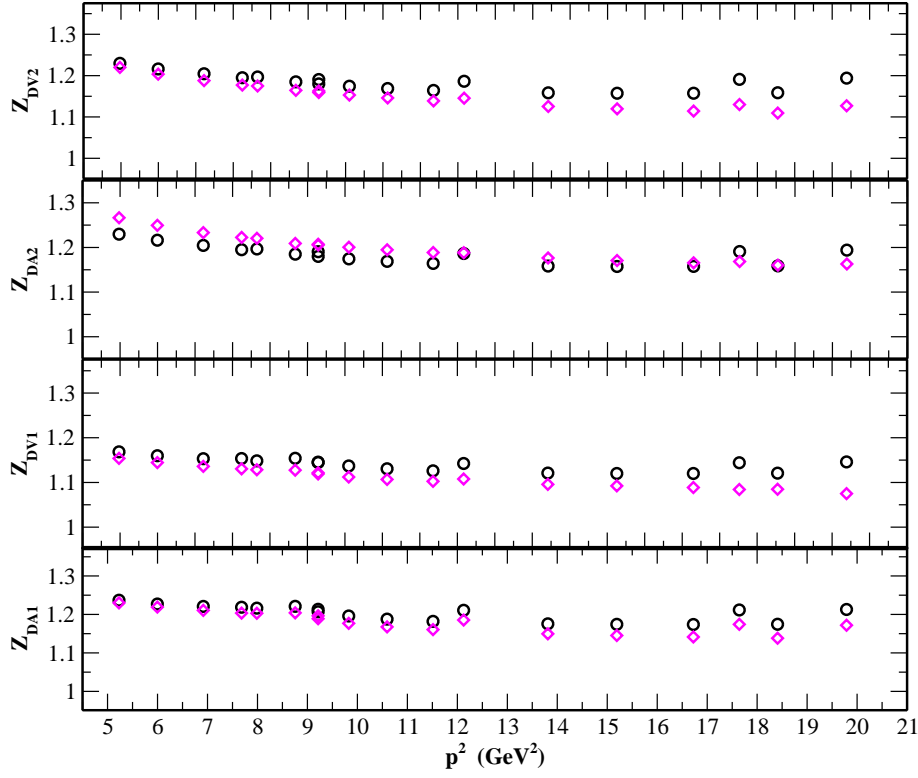}$\qquad\quad$}
\vspace{0.1cm}
   \caption{ Same as Fig~\ref{fig9}, but for $\beta=4.05$ and $\mu_0=0.008$.}
\label{fig10}
\end{figure}

\begin{figure}[h]
\centerline{\psfig{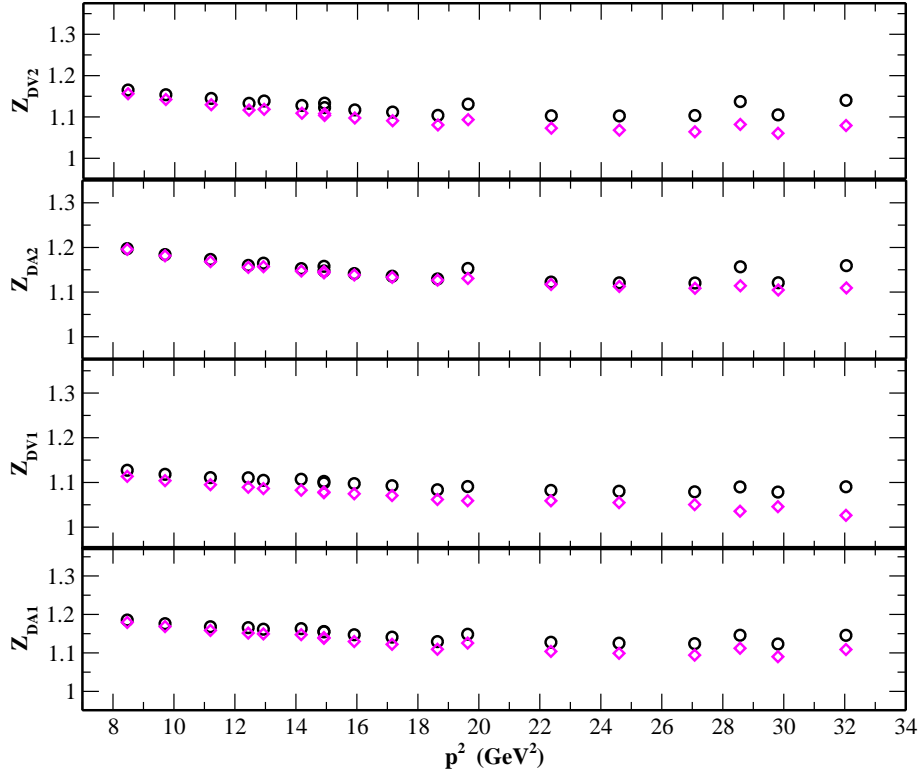}$\qquad\quad$}
\vspace{0.1cm}
   \caption{ Same as Fig~\ref{fig9}, but for $\beta=4.20$ and $\mu_0=0.0065$.}
\label{fig11}
\end{figure}

\subsection{RI$'$-MOM at a reference scale}
All our Z-factors have been evaluated for a range of renormalization
scales. In this subsection we use 2-loop perturbative expressions to
extrapolate to a scale $\mu = 1/a$ (the values for $a$ are taken from
Table~\ref{tab1}). Thus, each result is extrapolated to $1/a$, maintaining
the information of the initial renormalization scale at which it was
computed. Although the 3-loop formula is available for the following
expressions, the ${\cal O}(g^6)$ corrections are insignificant
compared to the lower order results.

The scale dependence is predicted by the renormalization group
(at fixed bare parameters), that is
\be
Z^{\rm RI'}_{\cal O}(\mu) = R_{\cal O}(\mu,\mu_0)\,Z^{\rm RI'}_{\cal O}(\mu_0)
\label{ZRI}
\ee
with
\be
\displaystyle
R_{\cal O}(\mu,\mu_0) =
   \left( \frac{\overline{g}^2(\mu^2)}{\overline{g}^2(\mu_0^2)} \right)
                  ^{\frac{\gamma^{\cal O}_0}{2 \beta_0}}
 \left( \frac{1 + \frac{\beta_1}{\beta_0} \frac{\overline{g}^2(\mu^2)}{16 \pi^2}}
        {1 + \frac{\beta_1}{\beta_0} \frac{\overline{g}^2(\mu_0^2)}{16 \pi^2}}
    \right)^{\frac{1}{2} \left( \frac{\gamma^{\cal O}_1}{\beta_1} - 
              \frac{\gamma^{\cal O}_0}{\beta_0} \right)}
\label{R}
\ee

\vspace{0.5cm}
\noindent
To 2 loops, the running coupling, $\beta$-function and
anomalous dimension $\gamma$ are as follows:
\bea
\label{g}
  \frac{\overline{g}^2 (\mu^2)}{16 \pi^2} &=& 
       \frac{1}{\beta_0 \ln(\mu^2/\Lambda^2)} -
        \frac{\beta_1}{\beta_0^3} \frac{\ln \ln(\mu^2/\Lambda^2) }
                              {\ln^2(\mu^2/\Lambda^2)} + \cdots   \\ [3ex]
  \beta_0 &=& 11 - \frac{2}{3} N_F \,,\, \beta_1 = 102 - \frac{38}{3} N_F  \\ [3ex]
  \gamma^{\cal O} (g) &=&  \gamma^{\cal O}_0 \frac{g^2}{16 \pi^2} + 
             \gamma^{\cal O}_1 \left( \frac{g^2}{16 \pi^2} \right)^2 + \cdots
\eea

The expressions for the anomalous dimension of the
fermion field and the twist-2 vector/axial operators are given in Ref.~\cite{Gracey:2003mr},
\bea
\gamma^{\mbox{\footnotesize{\rm RI$'$}}}_\psi(g) &=&  2\,\lambda C_F  \frac{g^2}{16\,\pi^2}  + 
2\,\Big[ \left( 9 \lambda^3 + 45 \lambda^2 + 223 \lambda + 225
  \right) C_A \nonumber \\
&&\phantom{\lambda C_F  \frac{g^2}{16\,\pi^2}  +}
- 54 C_F - \left( 80 \lambda + 72 \right) T_F N_F\Big] \frac{C_F}{36}
\left(\frac{g^2}{16\,\pi^2}\right)^2  \\[3ex]
\gamma^{\mbox{\footnotesize{\rm RI$'$}}}_{\overline\psi
  \gamma^{\{\mu}\,D^{\nu\}}\psi}(g) &=&
2\,\frac{8}{3}\,C_F\,\frac{g^2}{16\,\pi^2} + 
\frac{2}{54}\,C_F\,\Big[ \left( 27 \lambda^2 + 81 \lambda + 1434
  \right) C_A \nonumber \\
&&\phantom{\frac{8}{3}\,C_F\,\frac{g^2}{16\,\pi^2} + 
\frac{8}{54}\,C_F\,}
- 224 C_F - 504\,T_F N_F\Big]
\left(\frac{g^2}{16\,\pi^2}\right)^2  \quad,
\label{gRI}
\eea
where $T_F=1/2,\, C_A=N_c$. Using Eqs. (\ref{ZRI}) - (\ref{gRI}) we
obtain the Z-factors at $\mu=1/a$ for $\beta=3.9,\,4.05,$ and $4.20$,
which are plotted in Figs. \ref{fig9} - \ref{fig11}.

\subsection{Conversion to ${\overline{\rm MS}}$}

The passage to the continuum ${\overline{\rm MS}}$-scheme is
accomplished through use of a conversion factor which is computed up
to 3 loops in perturbation theory. By definition, this conversion
factor is the same for the vector and axial twist-2 renormalization
constant, but will differ for the cases $Z_{\rm DV1}\,(Z_{\rm DA1})$
and $Z_{\rm DV2}\,(Z_{\rm DA2})$, that is 
\bea
\label{CDV1}
C_{\rm DV1} \equiv C_{\rm DA1} =\frac{Z^{\overline{\rm MS}}_{\rm DV}}{Z^{\rm RI'}_{\rm DV1}}
\\[2ex]
\label{CDV2}
C_{\rm DV2} \equiv C_{\rm DA2} =\frac{Z^{\overline{\rm MS}}_{\rm DV}}{Z^{\rm RI'}_{\rm DV2}}\,.
\eea
This requirement for different conversion factors results from the
fact that the Z-factors in the continuum ${\overline{\rm MS}}$-scheme
do not depend on the external indices, $\mu,\,\nu$ (see Eq. (2.5) of
Ref.~\cite{Gracey:2003mr}), while the results in the RI$'$-MOM scheme
do depend on  $\mu$ and $\nu$. Of course the conversion factors take
a different value for each renormalization scale; actually, the
direction of the momentum is required to be known
(Eqs.~(\ref{conv1}) - (\ref{conv2})). 

\begin{figure}[h]
\centerline{\psfig{figure=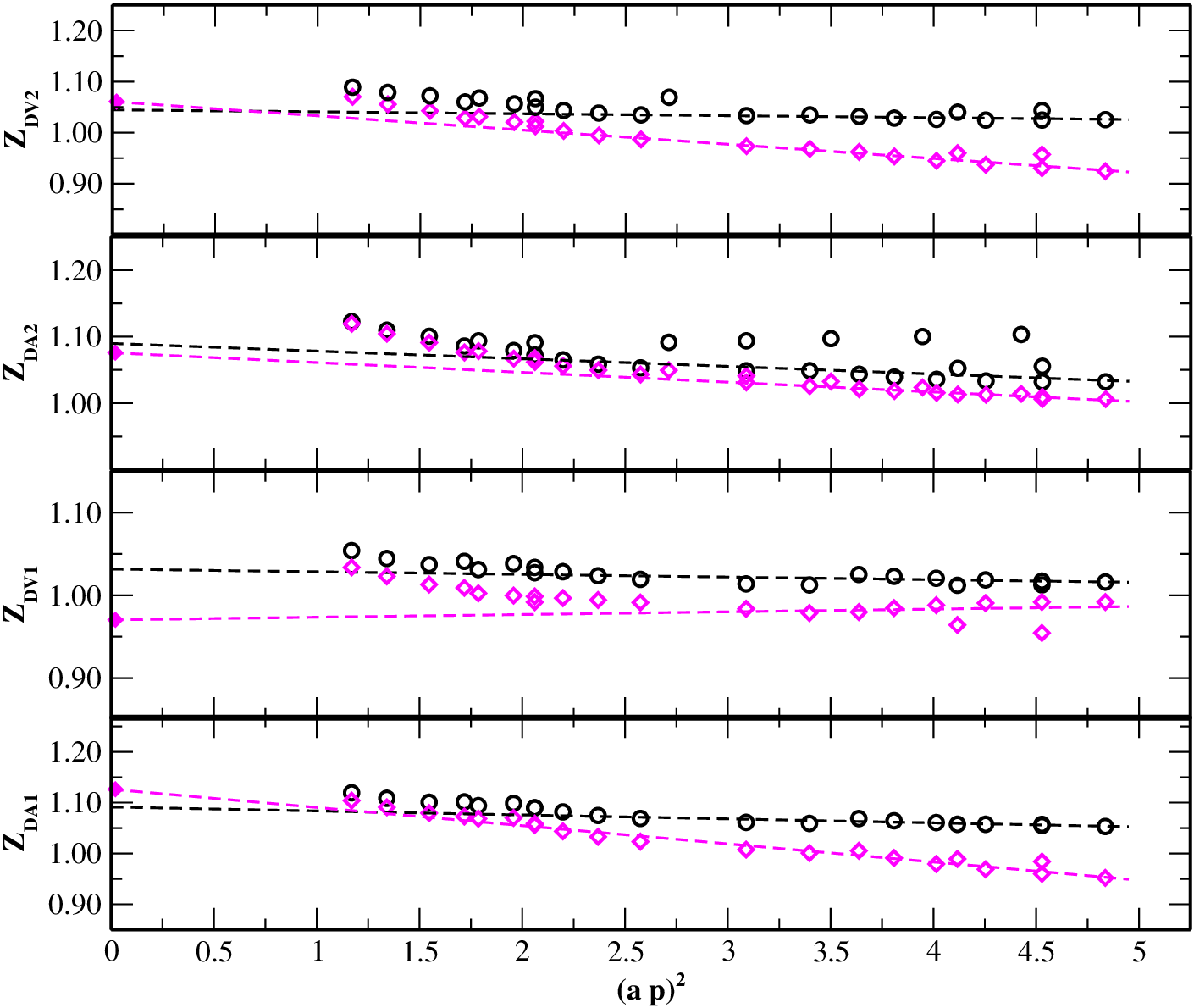,height=10.25truecm,clip=}$\qquad\quad$}
\vspace{0.1cm}
   \caption{ Renormalization factors at $\beta=3.9$, $\mu_0=0.0085$ in the ${\overline{\rm MS}}$-scheme
   at renormalization scale 2~GeV. Black circles correspond 
   to the unsubtracted results, while magenta diamonds correspond to the results with 
   perturbatively subtracted 1-loop $O(a^2)$ artifacts. 
   The lines show extrapolations to $a^2p^2=0$ using the subtracted results within
the range $a^2p^2=3-5$. }
\label{fig7}
\end{figure}

\begin{figure}[h]
\centerline{\psfig{figure=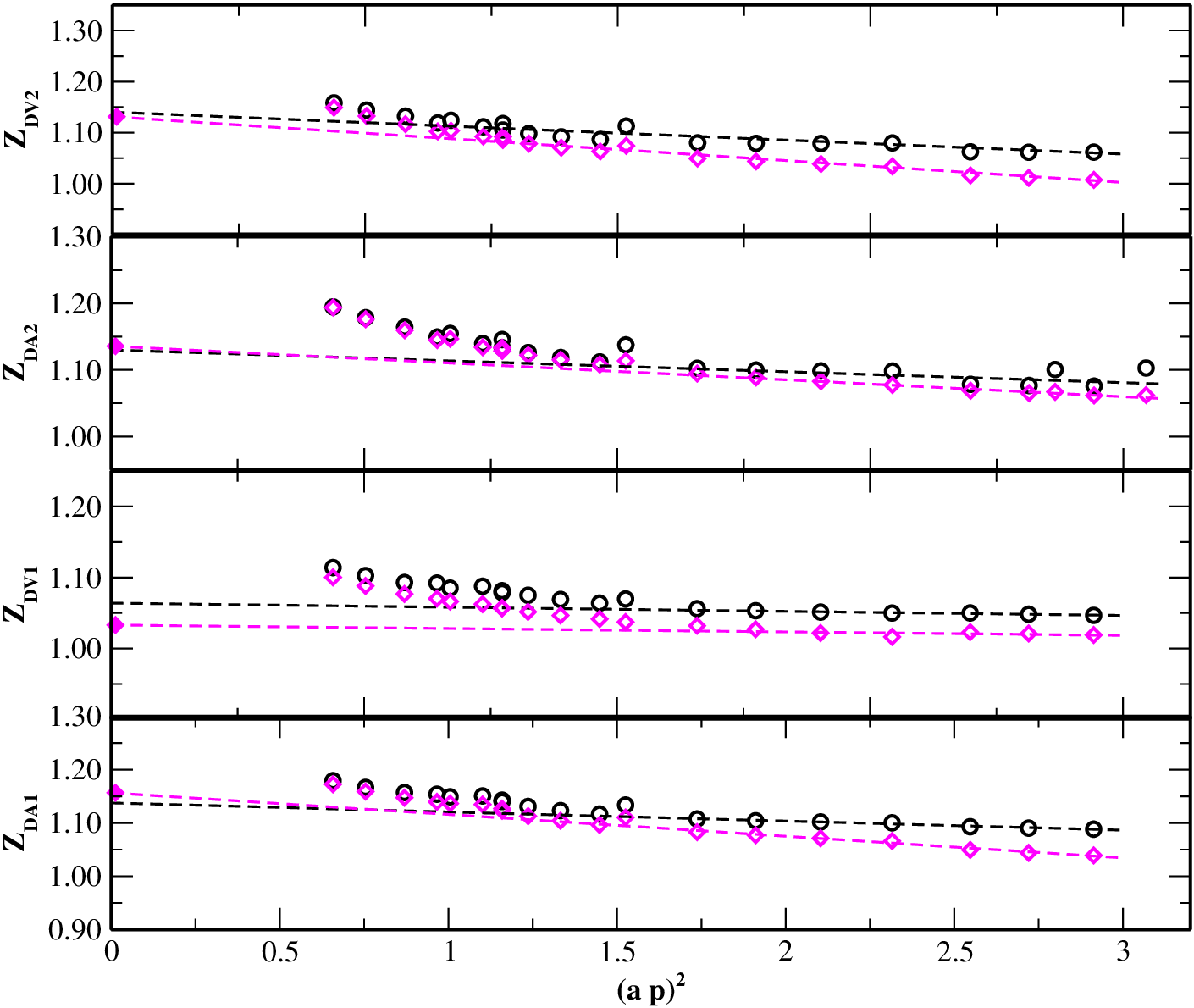,height=10.25truecm,clip=}$\qquad\quad$}
\vspace{0.1cm}
   \caption{ Renormalization factors at $\beta=4.05$, $\mu_0=0.008$ in the ${\overline{\rm MS}}$-scheme
   at renormalization scale 2~GeV. The black circles correspond 
   to the unsubtracted results, while the magenta diamonds to the results with 
   perturbatively subtracted one loop $O(a^2)$ artifacts. 
   The lines show extrapolations to $a^2p^2=0$ using the subtracted results within
the range $a^2p^2=1.9-3$.}
\label{fig12}
\end{figure}

\begin{figure}[h]
\centerline{\psfig{figure=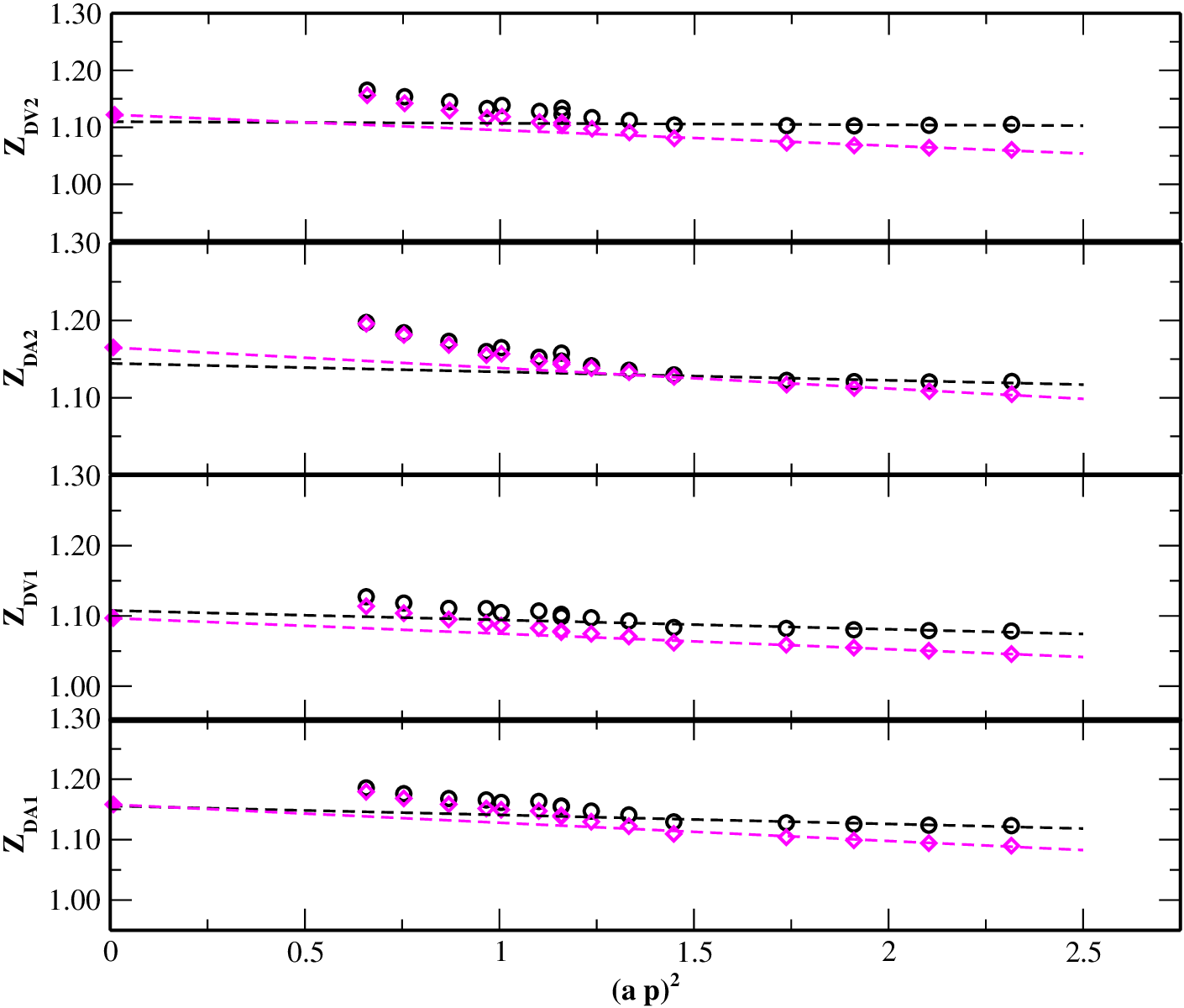,height=10.25truecm,clip=}$\qquad\quad$}
\vspace{0.1cm}
   \caption{ Renormalization factors at $\beta=4.20$, $\mu_0=0.0065$ in the ${\overline{\rm MS}}$-scheme
   at renormalization scale 2~GeV. The black circles correspond 
   to the unsubtracted results, while the magenta diamonds to the results with 
   perturbatively subtracted one loop $O(a^2)$ artifacts. 
   The lines show extrapolations to $a^2p^2=0$ using the subtracted results within
the range $a^2p^2=1.2-2.5$.}
\label{fig15}
\end{figure}

The 3-loop expressions for the conversion factors from our RI$'$-MOM
scheme (Eq.~(\ref{renormalization cond})) to the ${\overline{\rm MS}}$ do not appear
directly in the literature, but can be extracted using results from 
Ref.~\cite{Gracey:2003mr}. In the latter publication the reader can
find the conversion factor from an alternative definition of
RI$'$-MOM (which we denote by RI) to the usual ${\overline{\rm MS}}$,
$C_{\overline\psi \gamma^\mu D^\nu \psi}$. This alternative definition reads
\begin{equation}  
\left. \lim_{\epsilon \, \rightarrow \, 0} \left[ 
Z^{\mbox{\footnotesize{\rm RI}}}_\psi  
Z^{\mbox{\footnotesize{\rm RI}}}_{\overline\psi \gamma^\mu D^\nu \psi}  
\Sigma^{(1)}_{\overline\psi \gamma^\mu D^\nu \psi}(p) \right] \right|_{p^2 \, = \, 
\mu^2} ~=~ 1 ~,
\label{RI_Gracey}
\end{equation}  
where $\Sigma^{(1)}$ can be extracted from the bare amputated Green's function
as follows
\begin{eqnarray} 
G^{\mu\nu}_{\overline\psi \gamma^{\{\mu} D^{\nu\}} \psi} (p) &=&  
\langle \psi(p) ~ [ \overline\psi \gamma^{\{\mu} D^{\nu\}} \psi](0) ~ \overline\psi
(-p) \rangle \nonumber \\
&=& \Sigma^{(1)}_{\overline\psi \gamma^{\{\mu} D^{\nu\}} \psi}(p) \left( \gamma^\mu p^\nu
+ \gamma^\nu p^\mu - \frac{2}{d} \pslash \eta^{\mu\nu} \right) \nonumber \\
&& +~ \Sigma^{(2)}_{\overline\psi \gamma^{\{\mu} D^{\nu\}} \psi}(p) \frac{1}{p^2}\left( p^\mu p^\nu 
\pslash - \frac{p^2}{d} \pslash \eta^{\mu\nu} \right) \,.
\end{eqnarray} 

The author of Ref.~\cite{Gracey:2003mr} provides the 3-loop expression
for the renormalized $\Sigma^{(2)}$ in the scheme of Eq.~(\ref{RI_Gracey}) 
(note that by definition the renormalized $\Sigma^{(1)}$ equals 1 at $p^2=\mu^2$).
These elements can be used to reconstruct the renormalized Green's
function 
\begin{eqnarray} 
G^{\mu\nu\,,R}_{\overline\psi \gamma^{\{\mu} D^{\nu\}} \psi} (p)\Bigr|_{p^2=\mu^2} &=&  \Bigg[1\cdot\left( \gamma^\mu p^\nu
+ \gamma^\nu p^\mu - \frac{2}{d} \pslash \eta^{\mu\nu} \right) \nonumber \\
&& + \Sigma^{(2) ~ {\mbox{\footnotesize{\rm RI$^\prime$}}} ~
\mbox{\footnotesize{finite}}}_{\overline\psi 
\gamma^{\{\mu} D^{\nu\}} \psi}(p) \,
\frac{1}{p^2}\left( p^\mu p^\nu 
\pslash - \frac{p^2}{d} \pslash \eta^{\mu\nu} \right) \Bigg]_{p^2=\mu^2}\,,
\end{eqnarray} 
in which we apply our RI$'$-MOM condition in order to obtain
\be
\frac{Z^{\rm RI}_{\overline\psi \gamma^{\{\mu} D^{\nu\}} \psi}}{Z^{\rm{RI}'}_{\rm DV1}}\,\,,\quad \frac{Z^{\rm RI}_{\overline\psi \gamma^{\{\mu} D^{\nu\}} \psi}}{Z^{\rm{RI}'}_{\rm DV2}}
\ee
Once we have these two elements we extract the conversion factor of
Eqs.~(\ref{CDV1}) - (\ref{CDV2}) up to 3 loops,
\bea
C_{\rm DV1}(\mu)&=&\frac{Z^{\rm RI}_{\overline\psi \gamma^{\{\mu} D^{\nu\}} \psi}}{Z^{\rm{RI}'}_{\rm DV1}}\cdot 
\left(C_{\overline\psi \gamma^\mu D^\nu \psi}\right)^{-1} = 
\frac{Z^{\rm RI}_{\overline\psi \gamma^{\{\mu} D^{\nu\}} \psi}}{Z^{\rm{RI}'}_{\rm DV1}}\cdot\frac{Z^{\rm
    {\overline{\rm MS}}}_{\rm DV}}{Z^{\rm{RI}}_{\overline\psi \gamma^{\{\mu} D^{\nu\}} \psi}} \\
C_{\rm DV2}(\mu)&=&\frac{Z^{\rm RI}_{\overline\psi \gamma^{\{\mu} D^{\nu\}} \psi}}{Z^{\rm{RI}'}_{\rm DV2}}\cdot 
\left(C_{\overline\psi \gamma^\mu D^\nu \psi}\right)^{-1} = 
\frac{Z^{\rm RI}_{\overline\psi \gamma^{\{\mu} D^{\nu\}} \psi}}{Z^{\rm{RI}'}_{\rm DV2}}\cdot\frac{Z^{\rm
    {\overline{\rm MS}}}_{\rm DV}}{Z^{\rm{RI}}_{\overline\psi \gamma^{\{\mu} D^{\nu\}} \psi}}\,.
\eea
The conversion to the ${\overline{\rm MS}}$ is then given by
\bea
Z^{\overline{\rm MS}}_{\rm DV1}(\mu)=C_{\rm DV1}(\mu)\cdot Z^{\rm RI'}_{\rm DV1}(\mu)\\
Z^{\overline{\rm MS}}_{\rm DA1}(\mu)=C_{\rm DV1}(\mu)\cdot Z^{\rm RI'}_{\rm DA1}(\mu)
\eea
and
\bea
Z^{\overline{\rm MS}}_{\rm DV2}(\mu)=C_{\rm DV2}(\mu)\cdot Z^{\rm RI'}_{\rm DV2}(\mu)\\
Z^{\overline{\rm MS}}_{\rm DA2}(\mu)=C_{\rm DV2}(\mu)\cdot Z^{\rm RI'}_{\rm DA2}(\mu)\,,
\eea
which correspond to the Z-factors at the same renormalization scale in
the RI$'$. One wants to obtain the renormalization constants
at the scale of 2 GeV, and to do this we use the 2-loop formula in
Eq.~(\ref{R})-(\ref{g}) to evolve the scale from $\mu$ to 2 GeV. In
these formulas we need to insert the anomalous dimension in the 
${\overline{\rm MS}}$-scheme which read~\cite{Gracey:2003mr}
\bea
\gamma^{\mbox{\footnotesize{${\overline{\rm MS}}$}}}_{\overline\psi
  \gamma^{\{\mu}\,D^{\nu\}}\psi}(\alpha) = 2\,\frac{8}{3} C_F  \alpha
~+~ 2\,\frac{8 C_F}{27} \left[ 47 C_A - 14 C_F - 16 T_F N_F \right] \alpha^2
\eea
where $\alpha=g^2/(16 \pi^2)$. The additional factor of 2 that we included, comes from the different
definition of the anomalous dimension that leads to Ref.~\cite{Floratos:1977au}.
To summarize, the Z-factors in the continuum ${\overline{\rm MS}}$
-scheme at $\mu=2$ GeV are given by
\bea
Z^{\overline{\rm MS}}_{\rm DV1}(2 GeV)=R_{\rm DV}(2 GeV,\mu)\cdot C_{\rm DV1}(\mu)\cdot Z^{\rm RI'}_{\rm DV1}(\mu)\\[1.5ex]
Z^{\overline{\rm MS}}_{\rm DV2}(2 GeV)=R_{\rm DV}(2 GeV,\mu)\cdot C_{\rm DV2}(\mu)\cdot Z^{\rm RI'}_{\rm DV2}(\mu)\\[1.5ex]
Z^{\overline{\rm MS}}_{\rm DA1}(2 GeV)=R_{\rm DV}(2 GeV,\mu)\cdot C_{\rm DV1}(\mu)\cdot Z^{\rm RI'}_{\rm DA1}(\mu)\\[1.5ex]
Z^{\overline{\rm MS}}_{\rm DA2}(2 GeV)=R_{\rm DV}(2 GeV,\mu)\cdot C_{\rm DV2}(\mu)\cdot Z^{\rm RI'}_{\rm DA2}(\mu)\\[1.5ex]
\eea

\vspace{0.5cm}
For the $SU(N_c=3)$ colour group ($C_A=3$, $C_F=4/3$, $T_F=1/2$),
Landau gauge ($\lambda=0$), and general quark flavours, we have the
following conversion factors

\bea 
C_{\rm DV1} \equiv \frac{Z^{\overline{\rm MS}}_{\rm DV}}{Z^{\rm{RI}'}_{\rm DV1}} &=& 
1 + \alpha\,\Bigg[-\frac{136}{27} + \frac{64}{9}\,
    \frac{\mu_\mu^2-\frac{\mu_\mu^4}{\mu^2}}{\mu^2 + 8 \mu_\mu^2}\Bigg] \nonumber \\[1.5ex]
&+& 
 \alpha^2\,\Bigg[-\frac{128096}{729} + N_F\,\left(\frac{3208}{243} - 
     \frac{320}{9}\,
      \frac{\mu_\mu^2-\frac{\mu_\mu^4}{\mu^2}}{\mu^2 + 8 \mu_\mu^2}\right) + 
   \frac{248}{9}\,\zeta(3) + 
   \frac{\mu_\mu^2-\frac{\mu_\mu^4}{\mu^2}}{\mu^2 + 8 \mu_\mu^2}\,
    \Big(\frac{17792}{27} + \frac{320}{9}\,\zeta(3)\Big)\Bigg]\nonumber \\[1.5ex]
&+& 
 \alpha^3\,\Bigg[-\frac{627867571}{78732} - \frac{64\,\pi^4 }{729}+ 
   \frac{5588641}{2187}\,\zeta(3) + N_F^2\,\left(-\frac{149552}{6561} + 
     \frac{77440}{729}\,
      \frac{\mu_\mu^2-\frac{\mu_\mu^4}{\mu^2}}{\mu^2 + 8 \mu_\mu^2} - 
     \frac{256}{243}\,\zeta(3)\right) \qquad\nonumber \\[1.5ex]
&&\phantom{\alpha^3\,}
+ N_F\,\left(\frac{19947676}{19683} + 
     \frac{64\,\pi^4}{243} - \frac{1600}{27}\,\zeta(3) + 
     \frac{\mu_\mu^2-\frac{\mu_\mu^4}{\mu^2}}{\mu^2 + 8 \mu_\mu^2}\,
      \Big(-\frac{121024}{27} + \frac{9856}{81}\,\zeta(3)\Big)\right) \nonumber \\[1.5ex]
&&\phantom{\alpha^3\,}
- \frac{19420}{27}\,\zeta(5) + 
   \frac{\mu_\mu^2-\frac{\mu_\mu^4}{\mu^2}}{\mu^2 + 8 \mu_\mu^2}\,
    \Big(\frac{270701210}{6561} - \frac{2993992}{243}\,\zeta(3) + 
     \frac{349600}{81}\,\zeta(5)\Big)\Bigg]+ {\cal O}(\alpha^4)
\label{conv1}
\eea
%%%%%%%%%%%%%%%%%%%%%%%%%%%%%%%%%%%%%%%%%%%%%%%%%%%%%%%%%%%%%%%%%%%%%%%%%%%%%%%%%%%%%%%%%%%%%%%%%%%%%%%%%%%%%%%%%%%
\bea
%%%%%\label{conv2}
C_{\rm DV2} \equiv \frac{Z^{\overline{\rm MS}}_{\rm DV}}{Z^{\rm{RI}'}_{\rm DV1}} &=& 
1 + \alpha\,\Bigg[- \frac{124}{27}  -  \frac{16}{9} \,
     \frac{\mu_\mu^2 \mu_\nu^2}{\mu^2 (\mu_\mu^2 + \mu_\nu^2)} \Bigg] \nonumber \\[1.5ex]
&+& 
 \alpha^2\,\Bigg[- \frac{98072}{729}  + N_F\,\left( \frac{2668}{243}  + 
      \frac{80}{9} \, \frac{\mu_\mu^2 \mu_\nu^2}{\mu^2 (\mu_\mu^2 + \mu_\nu^2)} \right) + 
    \frac{268}{9} \,\zeta(3) +  \frac{\mu_\mu^2 \mu_\nu^2}{\mu^2 (\mu_\mu^2 + \mu_\nu^2)} \,
    \Big(- \frac{4448}{27}  -  \frac{80}{9} \,\zeta(3)\Big)\Bigg] \quad\nonumber \\[1.5ex]
&+& 
 \alpha^3\,\Bigg[- \frac{849683327}{157464}  -  \frac{64 \,\pi^4}{729} + 
    \frac{7809041}{4374} \,\zeta(3) + N_F^2\,\left(- \frac{105992}{6561}  - 
      \frac{19360}{729} \, \frac{\mu_\mu^2 \mu_\nu^2}{\mu^2 (\mu_\mu^2 + \mu_\nu^2)}  - 
      \frac{256}{243} \,\zeta(3)\right) \nonumber \\[1.5ex]
&&\phantom{\alpha^3\,}+ 
N_F\,\left( \frac{14433520}{19683}  + 
      \frac{64 \,\pi^4}{243} -  \frac{4184}{81} \,\zeta(3) + 
      \frac{\mu_\mu^2 \mu_\nu^2}{\mu^2 (\mu_\mu^2 + \mu_\nu^2)} \,
      \Big( \frac{30256}{27}  -  \frac{2464}{81} \,\zeta(3)\Big)\right) \nonumber \\[1.5ex]
&&\phantom{\alpha^3\,}
-\frac{36410}{81} \,\zeta(5) + 
    \frac{\mu_\mu^2 \mu_\nu^2}{\mu^2 (\mu_\mu^2 + \mu_\nu^2)} \,
    \Big(- \frac{135350605}{13122}  +  \frac{748498}{243} \,\zeta(3) - 
      \frac{87400}{81} \,\zeta(5)\Big)\Bigg]+ {\cal O}(\alpha^4)
\label{conv2}
\eea
 \noindent where $\alpha=g^2/(16 \pi^2)$ and $\zeta(n)$ is the Riemann Zeta function.

A ``renormalization window'' should exist for $\Lambda_{QCD}^2 <<
\mu^2 << 1/a^2$ where perturbation theory holds and finite-$a$
artifacts are small, leading to scale-independent results
(plateau). In practice such a condition is hard to satisfy: The right
inequality is extended to $(2-5)/a^2$ leading to lattice artifacts in
our results that are of ${\cal O}(a^2p^2)$. Fortunately our
perturbative calculations allow us to subtract the leading
perturbative $O(a^2)$ lattice artifacts which alleviates the
problem. To remove the remaining $O(a^2p^2)$ artifacts we extrapolate
linearly to $a^2p^2=0$ as demonstrated in Figs. \ref{fig7} -\ref{fig15}.
The statistical errors are negligible and therefore an estimate of the
systematic errors is important. We note that, in general, the evaluation
of systematic errors is difficult. The largest systematic error comes
from the choice of the momentum range to
use for the extrapolation to $a^2p^2=0$. One way to estimate
this systematic error is to vary the momentum range where we perform
the fit. Another approach is to fix a range and then eliminate a given
momentum in the fit range and refit. The spread of the results about
the mean gives an estimate of the systematic error. In the final
results we give as systematic error the largest one from using these
two procedures which is the one obtained by modifying the fit
range.We choose the same momentum
range in physical units for all $\beta$-values
and we thus extract all renormalization constants
using the same physical momentum range, $p^2\sim 15-32$ (GeV)$^2$. This
momentum range has been chosen so that we are in a region where
an approximate
plateau is seen at each $\beta$. We also note that the ${\cal O}(a^2)$ perturbative terms
which we subtract, decrease as $\beta$ increases, as expected and the
values extracted from subtracted and unsubtracted data agree
when extrapolated to $a=0$. 
The momentum range in lattice units at each $\beta$ is
 as follows: $\beta=3.9: a^2p^2\sim 3-5$, $\beta=4.05:
a^2p^2\sim 1.9-3$, $\beta=4.20: a^2p^2\sim 1.2-2.5$ and   as can be seen in Figs. \ref{fig7}
-\ref{fig15} within these ranges
 the data fall on a straight line of a small slope.

\noindent Our final results for the $Z$-factors in the $\overline{\rm MS}$-scheme at 2~GeV are
given in Table~\ref{tab7}, which have been obtained by extrapolating
linearly in $a^2p^2$, using the fixed momentum range $p^2\sim 15-32$ (GeV)$^2$.

\begin{table}[h]
\begin{center}
\begin{minipage}{15cm}
\begin{tabular}{lr@{}lr@{}lr@{}lr@{}l}
\hline
\hline\\[-2.25ex]
\multicolumn{1}{c}{$\beta$}&
\multicolumn{2}{c}{$\,\,Z_{\rm DV1}\,\,$} &
\multicolumn{2}{c}{$\,\,Z_{\rm DV2}\,\,$} &
\multicolumn{2}{c}{$\,\,Z_{\rm DA1}\,\,$} &
\multicolumn{2}{c}{$\,\,Z_{\rm DA2}\,\,$}  \\
\hline
\hline\\[-2.25ex]
3.90  &$\,\,$&0.970(34)(26) &$\,\,$&1.061(23)(29)   &$\,\,$&1.126(22)(78)    &$\,\,$&1.076(5)(1) \\ 	 				       
4.05  &$\,\,$&1.033(11)(14) &$\,\,$&1.131(23)(18)   &$\,\,$&1.157(9)(7)      &$\,\,$&1.136(5)     \\         	   			       
4.20  &$\,\,$&1.097(4)(6)   &$\,\,$&1.122(7)(10)    &$\,\,$&1.158(7)(7)      &$\,\,$&1.165(5)(10) \\
\hline
\hline
\end{tabular}
\end{minipage}
\end{center}
\caption{Renormalization constants $Z_{\rm DV}$ and $Z_{\rm DA}$ in
  the ${\overline{\rm MS}}$ scheme. The above values have been
  obtained by extrapolating linearly in $a^2p^2$. Statistical errors are
  are shown in the first parenthesis. The error in the
  second parenthesis is the systematic error due to the extrapolation,
  namely the difference between results using the fit range
  $p^2\sim 15-32$ (GeV)$^2$ and the range $p^2\sim
  17-24$ (GeV)$^2$. An error  smaller than
the last digit given for the mean value is not quoted.} 
\label{tab7}
\end{table}

\section{Conclusions}
\label{sec6}
The values of the renormalization factors for the one-derivative twist-2 operators are
calculated non-perturbatively. The method of choice is to use a momentum dependent source and extract the renormalization constants for all the relevant operators. 
This leads to a very accurate evaluation of these renormalization  factors using a 
small ensemble of gauge configurations. 
The accuracy of the results allows us to check for any light quark mass dependence.
For all the renormalization constants studied in this work we do not find
 any light quark mass dependence within our small statistical errors. 
Therefore it suffices to calculate them
at a given quark mass.
 We also show that, despite of using lattice spacing smaller
than 1~fm, ${\cal O}(a^2)$ effects are sizable. We perform a perturbative
subtraction of ${\cal O}(a^2)$  terms. This leads to a smoother dependence
of the renormalization constants on the momentum values at which they are  extracted.
 Residual
${\cal O}(a^2p^2)$ effects are removed by extrapolating to zero.
In this way we can accurately determine the renormalization constants
 in the RI$^\prime$-MOM scheme. In order to compare with experiment we
convert our values to the $\rm \overline{MS}$ scheme at a scale of 2~GeV.
The statistical errors are in general smaller than the systematic. The
latter are estimated by changing the window of values of the momentum used
to extrapolate to  $a^2p^2=0$. Our final values are given in Table~\ref{tab7}.

\section{Acknowledgments}
This work was partly supported by funding received from the
Cyprus Research Promotion Foundation under contracts EPYAN/0506/08,
and TECHNOLOGY/$\Theta$E$\Pi$I$\Sigma$/0308(BE)/17.

\bibliographystyle{apsrev}                     % Style for bibliography
\bibliography{gpdref}

\end{document}